\documentclass[12pt,a4paper]{article}
\usepackage{a4wide}
\usepackage{amsmath}
\usepackage{amssymb}
\usepackage{epsfig}
\usepackage{subfigure}
\usepackage{exscale}
\usepackage{float}
\usepackage{bbm}
\usepackage[numbers,sort&compress]{natbib}
\usepackage{pst-plot, pstricks,pst-math}
\usepackage{fancybox,amssymb,color}
\usepackage{graphicx}
\usepackage{pstricks, color, graphicx, epsfig, psfrag}
\usepackage{amsfonts,amsmath,amssymb,slashed}
\usepackage{dsfont}
\usepackage{bbm,bm}
\usepackage{fancyhdr, a4wide}
\usepackage[english]{babel}
\usepackage{subfigure}

\makeatletter
\@addtoreset{equation}{section}
\makeatother

\title{Quantum Chromodynamics, Antiferromagnets and XY Models from a Unified Point of View}

\author{Christoph P.\ Hofmann$^a$ \\ \\
\normalsize{$^a$ Facultad de Ciencias, Universidad de Colima} \\
\vspace{0.3cm}
\normalsize {Bernal D\'iaz del Castillo 340, Colima C.P.\ 28045, Mexico} \\}

\begin{document}

\maketitle

\begin{abstract} \normalsize

Antiferromagnets and quantum XY magnets in three space dimensions are described by an effective Lagrangian that exhibits the same
structure as the effective Lagrangian of quantum chromodynamics with two light flavors. These systems all share a spontaneously broken
internal symmetry O($N$) $\to$ O($N$-1). Although the respective scales differ by many orders of magnitude, the general structure of the
low-temperature expansion of the partition function is the same. In the nonabelian case, logarithmic terms of the form $T^8 \ln T$ emerge
at three-loop order, while for $N$=2 the series only involves powers of $T^2$. The manifestation of the Goldstone boson interaction in the
pressure, order parameter, and susceptibility is explored in presence of an external field.

\end{abstract}


\maketitle

\section{Motivation}
\label{Intro}

The systematic effective Lagrangian method is well-established in particle physics. The prime example is chiral perturbation theory -- the
low-energy effective field theory of quantum chromodynamics (QCD). The same techniques can be applied to condensed matter systems whenever
their low-energy physics is dominated by Goldstone bosons. In the present study, we consider $d$=3+1 dimensional systems characterized by
a spontaneously broken internal symmetry O($N$) $\to$ O($N$-1). This includes collective quantum magnetism: antiferromagnets ($N$=3) and
quantum XY magnets ($N$=2) -- but it also covers particle physics, namely two-flavor QCD ($N$=4). We present the evaluation of the
partition function up to three-loop order and discuss the impact of the Goldstone boson interaction at low temperatures and in presence of
an explicit symmetry breaking parameter (magnetic field, staggered field, nonzero quark mass). 

Whereas the microscopic Hamiltonians of the quantum XY model and the Heisenberg antiferromagnet are invariant under O(2) and O(3),
respectively, their ground states with antialigned spins, are not -- the staggered magnetization vector, i.e. the order parameter, is
different from zero. The spontaneously broken symmetry is only approximate because the staggered field explicitly breaks O(2) and O(3).
Accordingly, the magnons or spin-waves that represent the Goldstone bosons, develop an energy gap. The case O(4) $\to$ O(3) is locally
isomorphic to $SU(2) \times SU(2) \to SU(2)$ and hence describes quantum chromodynamics with a massless up- and down-quark exhibiting
a spontaneously broken chiral symmetry. In nature, chiral symmetry is only approximate: the quark masses are nonzero and therefore the
Goldstone bosons -- the pions -- have a small mass.

Appreciating the universal nature of the effective field theory method, we identify differences and analogies between systems as disparate
as quantum magnets and quantum chromodynamics. The structure of the low-temperature expansion is an immediate consequence of the
spontaneously broken symmetry. Also the question whether or not logarithmic terms of the form $T^n \ln T$ appear in the low-temperature
series, can be answered from a universal point of view. The specific nature of the system only reflects itself in the actual numerical
values of a few low-energy effective constants.

In the free energy density, the first correction to the leading free Bose gas term ($\propto T^4$) is of order $T^6$. As we demonstrate, a
logarithmic term $T^8 \ln T$ emerges at next-to-next-to-leading order (NNLO), provided that we are dealing with a nonabelian symmetry
($N \ge 3$). Likewise, in the order parameter and the susceptibility, logarithmic contributions show up. On the other hand, in the abelian
case, the temperature series are characterized by simple powers of $T^2$.

The Goldstone boson interaction in the pressure -- depending on the strength of the external field and the specific $N$ -- may be
repulsive, attractive, or zero at low temperatures. Regarding the order parameter and the susceptibility, counterintuitive effects may
occur: the $T$-dependent interaction contribution in the order parameter (susceptibility) may become positive (negative). It should be
noted that in QCD where the quark masses ("external field") are fixed at their nonzero values, no such counterintuitive effects occur at
low temperatures in the physical region where the pion mass is $M_{\pi} \approx 139.6 MeV$: going from $T$=0 to finite temperature, the
$T$-dependent interaction contribution in the quark condensate (susceptibility) is negative (positive).

Before the systematic effective field theory analysis of the $d$=3+1 quantum XY model \citep{Hof16b}, the properties of this system at low
temperatures where the spin waves dominate the physics, were only known to one-loop accuracy (see, e.g., Refs.~\citep{BEL70,OB78,UTT79,
AHN80,AHN81,WOH91,OHW94,GP95,SFBR01,CEHMS02}). In the case of the Heisenberg antiferromagnet, the low-temperature series for the staggered
magnetization has been reported in the pioneering article by Oguchi \citep{Ogu60}, but he apparently missed -- much like subsequent
investigators \citep{KL61,OH63,ABK61a,ABK61b,Har64,AC64,Liu66,NT69,CS70,FR70,Gho73} -- logarithmic terms $T^n \ln T$ in the
low-temperature series. This demonstrates the power of the effective Lagrangian method in the condensed matter domain: in the present
applications it proves to be superior to standard methods like spin-wave theory. On the other hand, the properties of quantum
chromodynamics at finite temperatures have been the subject of many publications (see, e.g., Refs.~\citep{GL89,SV96,Smi97,NK10,GT11,TG12,
AokEA09,BazEA12a,BazEA12b,BucEA14,BhaEA14,BazEA14,BorEA15}. Still, some additional material is presented here that is new to the best of
our knowledge.

The article is organized as follows. The microscopic -- or underlying -- theories, as well as the corresponding effective Lagrangians, are
briefly discussed in Sec.~\ref{MicroEff}. The results for general $N$ are presented in Sec.~\ref{generalN}, where we focus on the
structure of the low-temperature expansion for the free energy density. In Sec.~\ref{LowTSeries}, the low-temperature series for the
pressure, order parameter and susceptibility are considered. The impact of the Goldstone-boson interaction in these observables is
illustrated in various figures. In Sec.~\ref{conclusions} we finally present our conclusions. Technical details of the calculation are
contained in an appendix.

\section{Underlying Versus Effective Description}
\label{MicroEff}

Chiral perturbation theory has been transferred to condensed matter systems in Ref.~\citep{Leu94a}, and the formalism has been generalized
in Ref.~\citep{ABHV14}. Here we focus on finite-temperature effective field theory -- outlines going beyond the brief sketch presented in
this section, can be found in Ref.~\citep{Hof16b} and the references provided there.

The microscopic Hamiltonian for the antiferromagnetic quantum XY model in $d$=3+1 is given by
\begin{equation}
{\cal H} = - J \sum_{\langle xy \rangle} (S^1_x S^1_y + S^2_x S^2_y) - {\vec H_s} \cdot \sum_x (-1)^{\frac{x_1+x_2+x_3}{\hat a}} {\vec S_x} \, ,
\qquad J < 0 \, .
\end{equation}
The summation $\langle xy \rangle$ extends over all pairs of nearest neighbors on a simple cubic lattice, separated by the distance
$\hat a$. The quantity $J$ is the exchange integral, and ${\vec H_s} = (H_s, 0)$ is the staggered field pointing into the 1-direction in
the XY plane. The model can be extended by including the third spin component,
\begin{eqnarray}
{\cal H} & = & - J \sum_{\langle xy \rangle} (S^1_x S^1_y + S^2_x S^2_y + S^3_x S^3_y) - {\vec H_s} \cdot \sum_x (-1)^{\frac{x_1+x_2+x_3}{\hat a}}
{\vec S_x} \, , \qquad J < 0 \, , \nonumber \\
& & {\vec H_s} = (H_s, 0, 0) \, ,
\end{eqnarray}
which defines the Heisenberg antiferromagnet in a staggered field. It should be noted that on a bipartite lattice -- and only for the
abelian case $N$=2 -- there is a one-to-one correspondence between the XY quantum antiferromagnet in a staggered field and the
ferromagnetic ($J > 0$) quantum XY model in a magnetic field. 

At zero temperature and infinite volume, the staggered magnetization $\Sigma_s$ and the spontaneous magnetization $\Sigma$ are different
from zero,
\begin{eqnarray}
\Sigma_s & = & \langle 0 | \sum_x (-1)^{(x_1+x_2+x_3)/{\hat a}} S^1_x | 0 \rangle / V \, , \qquad (N = 2,3 ) \nonumber \\
\Sigma & = & \langle 0 | \sum_x S^1_x | 0 \rangle / V \, , \qquad (N = 2)
\end{eqnarray}
indicating that the O($N$) symmetry is spontaneously broken. As a consequence, $N$-1 Goldstone bosons emerge -- these are the well-known
spin waves or magnons.

We now leave quantum magnetism and consider the strong interaction. The QCD Lagrangian with two light quarks is given by
\begin{equation}
{\cal L}_{QCD} = - \frac{1}{2g^2} \mbox{tr}_cG_{\mu\nu} G^{\mu\nu} + \bar{q} i \gamma^\mu D_\mu q - \bar{q}\, m\,q \, , \qquad (q = u, d) \, ,
\end{equation}
where all quantities have the usual meaning (see e.g. Ref.~\citep{Leu95}).\footnote{In the following we refer to the {\it isospin limit}
where the two light quark masses are identical: $m_u = m_d$.} The quark condensate
\begin{equation}
\langle 0 | \, {\bar u} u \, | 0 \rangle = \langle 0 | \, {\bar d} d \, | 0 \rangle
\end{equation}
represents the order parameter, its nonzero value signaling that the chiral symmetry $SU(2) \times SU(2) \approx O(4)$ is spontaneously
broken. The three pions are the Goldstone bosons, originating from the spontaneously broken chiral symmetry.

The low-energy physics of all three systems -- XY models, Heisenberg antiferromagnets and quantum chromodynamics -- is thus dominated by
the corresponding Goldstone bosons, such that the effective Lagrangian method ("chiral perturbation theory" in the context of QCD) can be
applied. In the following we assume that we are dealing with a spontaneously broken symmetry O($N$) $\to$ O($N$-1), keeping in mind that
the cases $N=\{2,3,4\}$ refer to the physical realizations of interest.

The Goldstone boson physics is universal: the specific terms in the effective Lagrangian are a consequence of the symmetries of the
underlying (microscopic) theory. A particular physical realization then corresponds to a specific numerical set of low-energy effective
constants. Whereas the respective scales involved may differ by many orders of magnitude, the structure of the low-energy
(low-temperature) expansion is universal.

The leading term in the effective Lagrangian that describes the symmetry breaking pattern O($N$) $\to$ O($N$-1), is of order $p^2$ as it
contains two derivatives,
\begin{equation}
\label{L2}
{\cal L}^2_{eff} = \mbox{$ \frac{1}{2}$} F^2 \partial_{\mu} U^i \partial^{\mu} U^i + \Sigma_s H_s^i U^i \, .
\end{equation}
The $N$-1 Goldstone bosons that dominate the low-energy physics are contained in the unit vector $U^i$
\begin{equation}
U^i = (U^0, U^a) \, , \qquad U^0 = \sqrt{1 - U^a U^a} \, , \qquad a=1,2, \dots, N-1 \, .
\end{equation}
Note that we are actually dealing with (pseudo-)Goldstone bosons: in presence of an external field $H_s$, the Goldstone bosons become
massive
\begin{equation}
\omega = \sqrt{v^2 {\vec k}^2 + v^4 M^2} \, .
\end{equation}
In the context of magnetic systems, $v$ is the spin-wave velocity that we set to one. The relation between the Goldstone boson mass and
the external field is given by
\begin{equation}
\label{GBMass}
M^2 = \frac{{\Sigma}_s H_s}{F^2} \, .
\end{equation}
At leading order we have two low-energy effective constants: $F$ ("pion decay constant" in QCD) and $\Sigma_s$ (order parameter at zero
temperature and infinite volume). The next-to-leading order effective Lagrangian is of order $p^4$, 
\begin{eqnarray}
\label{Leff4}
{\cal L}^4_{eff} & = & e_1 (\partial_{\mu} U^i \partial^{\mu} U^i)^2 + e_2 (\partial_{\mu} U^i \partial^{\nu} U^i)^2
+ k_1 \frac{\Sigma_s}{F^2} (H_s^i U^i) (\partial_{\mu} U^k \partial^{\mu} U^k) \nonumber \\
& & + k_2 \frac{{\Sigma}_s^2}{F^4} (H_s^i U^i)^2 + k_3 \frac{{\Sigma}_s^2}{F^4} H_s^i H_s^i \, ,
\end{eqnarray}
and contains terms with up to four derivatives and up to two powers of the external field $H_s$ that counts as order $p^2$. The
interpretation of the external field depends on context: magnetic field ($N$=2), staggered field ($N=2,3$), nonzero quark mass
($N$=4). Note that ${\cal L}^4_{eff}$ involves five next-to-leading order (NLO) low-energy effective constants. In QCD, their numerical
values are known -- in connection with magnetic systems we have to estimate their size (see below).

While the QCD Lagrangian and the corresponding effective Lagrangian are Lorentz-invariant by construction, it is still legitimate to
maintain a (pseudo-)Lorentz-invariant framework in connection with quantum magnetism on the effective level. First, the leading-order
effective Lagrangian is strictly Lorentz-invariant (where $c = v \equiv 1$ is the spin-wave velocity): this is an accidental symmetry.
Anisotropies start manifesting themselves at order $p^4$: indeed, in ${\cal L}^4_{eff}$ one should take into account all terms that are
allowed by, e.g., the cubic lattice geometry. However, the interaction part of the low-temperature expansion of the partition function is
only affected at next-to-next-to leading order by these additional terms (see below), such that our conclusions are not altered. This is
why it is justified to write ${\cal L}^4_{eff}$ in a (pseudo-)Lorentz-invariant form.

\section{Results for the General Case O($N$) $\to$ O($N$-1)}
\label{generalN}

While the $d$=3+1 quantum XY model ($N$=2) has been discussed in Ref.~\citep{Hof16b}, in the present study we focus on the nonabelian
case $N \ge 3$ that includes antiferromagnets and quantum chromodynamics. The main observable of interest is the free energy density, from
which the pressure, order parameter, and susceptibility can be derived. Details of the three-loop evaluation of the partition function can
be found in the appendices of Refs.~\citep{Hof16b,Hof99b}. In addition, pedagogic outlines on the effective Lagrangian method are provided
by Refs.~\citep{Leu95,Sch03,Bra10}.

\begin{figure}
\begin{center}
\includegraphics[width=15cm]{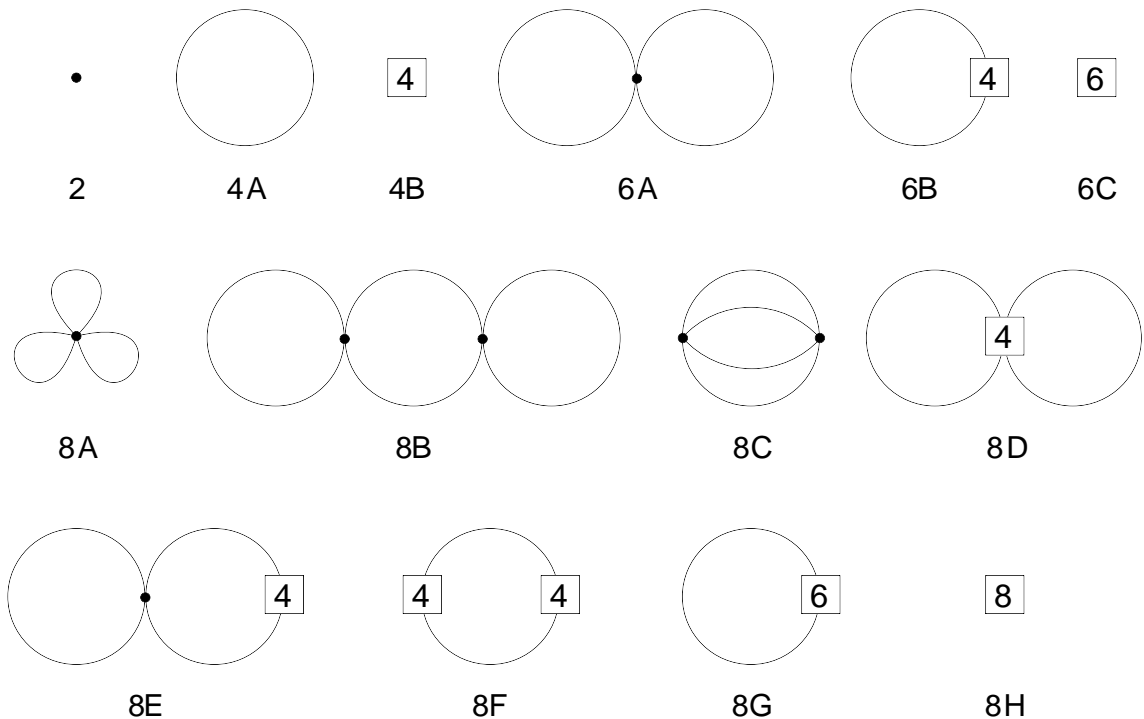}
\end{center}
\caption{Feynman diagrams: Low-temperature expansion of the partition function up to three-loop order $T^8$ for $d$=3+1 dimensional systems
with a spontaneously broken symmetry O($N$) $\to$ O($N$-1). Filled circles correspond to ${\cal L}^2_{eff}$ -- vertices related to the
higher-order pieces ${\cal L}^4_{eff},{\cal L}^6_{eff}, {\cal L}^8_{eff}$ of the effective Lagrangian are denoted by the numbers $4,6,8$.
Each loop is suppressed by $T^2$.}
\label{figure1}
\end{figure}

The Feynman diagrams needed up to three-loop order $p^8$, are depicted in Fig.~\ref{figure1}. Their evaluation leads to the following
expression for the free energy density at low temperatures, valid for (pseudo-)Lorentz-invariant systems with a spontaneously broken
internal symmetry O($N$) $\to$ O($N$-1):
\begin{equation}
\label{freeED}
z = z_0 - \mbox{$ \frac{1}{2}$} (N-1) g_0 - 4 \pi a (g_1)^2 - \pi g \, \Big[ b - \frac{j}{{\pi}^3 F^4} \Big] + {\cal O} (p^{10}) \, .
\end{equation}
The individual quantities have the following meaning:\footnote{In parenthesis we indicate where in appendix A of Ref.~\citep{Hof16b} the
respective definitions can be found.} free energy density at $T$=0 ($z_0$, Eq.~(A.28)), kinematical functions ($g_0, g_1, g$, Eqs.~(A.25)
and (A.41)). On the other hand, for the parameters $a, b$, and the three-loop function $j$, the explicit expressions are provided below,
as they all depend on $N$ which leads to some subtle effects.

The functions $g_0, g_1, g$ and $j$ depend on the dimensionless ratio $\sigma$ (or $\tau$),
\begin{equation}
\sigma = \frac{M_{\pi}}{2 \pi T} , \qquad \tau = \frac{T}{M_{\pi}} ,
\end{equation}
where $M_{\pi}$ is the Goldstone boson mass,
\begin{equation}
\label{renMass}
M_{\pi}^2 = \frac{{\Sigma}_s H_s}{F^2} + \Big[ 2 (k_2 - k_1) + (N-3) \, \lambda \Big] \frac{({\Sigma}_s H_s)^2}{F^6}
+ c \frac{({\Sigma}_s H_s)^3}{F^{10}} + {\cal O} (H_s^4) \, .
\end{equation}
The parameters $a$ and $b$ in Eq.~(\ref{freeED}) involve NLO effective constants and read
\begin{eqnarray}
\label{ConstAB}
a & = & - \frac{(N-1)(N-3)}{32{\pi}} \frac{{\Sigma}_s H_s}{F^4} + \frac{N-1}{4{\pi}} \frac{({\Sigma}_s H_s)^2}{F^8} \,
\Bigg\{ \Big[(N+1)(e_1 + e_2) + k_2 - k_1 \Big] \nonumber\\
& & - \frac{(N-1)^2}{2} \, \lambda - \frac{3N^2 + 32N - 67}{768 {\pi}^2} \Bigg\} \, , \nonumber \\ 
b & = & \frac{N-1}{{\pi}F^4} \, \Bigg\{ \Big[ 2 e_1 + N e_2 \Big] - \frac{5(N-2)}{3} \, \lambda - \frac{N-2}{16{\pi}^2} \Bigg\} \, .
\end{eqnarray}
Note that $\lambda$,
\begin{eqnarray}
\label{lambda}
\lambda & = & \mbox{$ \frac{1}{2}$} \, (4 \pi)^{-d/2} \, \Gamma(1-{\mbox{$ \frac{1}{2}$}}d) M^{d-4} \nonumber\\
& = & \frac{M^{d-4}}{16{\pi}^2} \, \Bigg[ \frac{1}{d-4} - \mbox{$ \frac{1}{2}$} \{ \ln{4{\pi}} + {\Gamma}'(1) + 1 \}
+ {\cal O}(d-4) \Bigg] \, ,
\end{eqnarray}
is singular in the limit $d \to 4$. The point is that this infinity can be absorbed by the NLO effective constants that show up in
$M_{\pi}^2, a, b$. Whether $e_1, e_2, k_1, k_2$ get renormalized logarithmically or do not require renormalization, depends on $N$.
Inspecting the $N$-dependence of $b$ and $M_{\pi}^2$, one notices the following subtleties. The parameter $\lambda$ does not appear in $b$
if $N$=2 -- therefore, in this case, the combination $e_1 + e_2$ of NLO effective constants is finite and does not require logarithmic
renormalization. Similarly, for $N$=3, the second term in the renormalized mass $M_{\pi}^2$ is free of $\lambda$ -- here the combination
$k_2 - k_1$ is finite and no logarithmic renormalization is needed. In all other cases (including $a$), the NLO effective constants are
renormalized logarithmically. The explicit renormalized expressions for $M_{\pi}^2, a$ and $b$, along with more details on the
renormalization procedure, can be found in the appendix.

\begin{figure}
\begin{center}
\includegraphics[width=14cm]{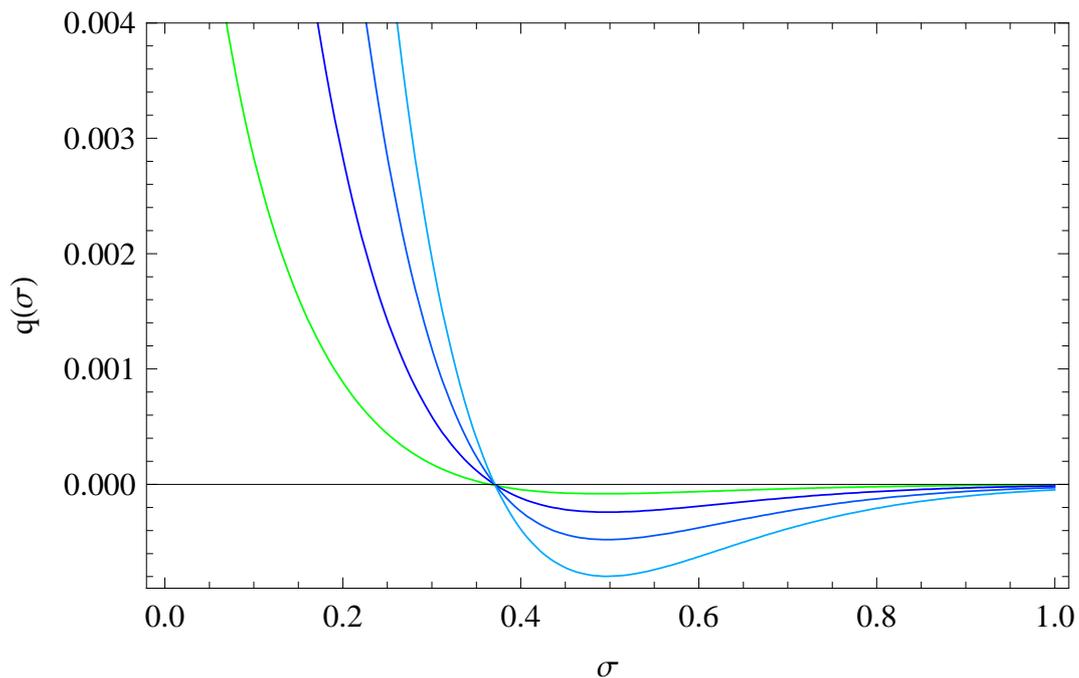}
\end{center}
\caption{[Color online] The three-loop function $q(\sigma)$ in terms of the parameter $\sigma = M_{\pi}/2\pi T$ for $N = \{ 3,4,5,6 \}$
from top to bottom in the figure (vertical cut at $\sigma = 0.5$).}
\label{figure2}
\end{figure}

\begin{figure}
\begin{center}
\includegraphics[width=14cm]{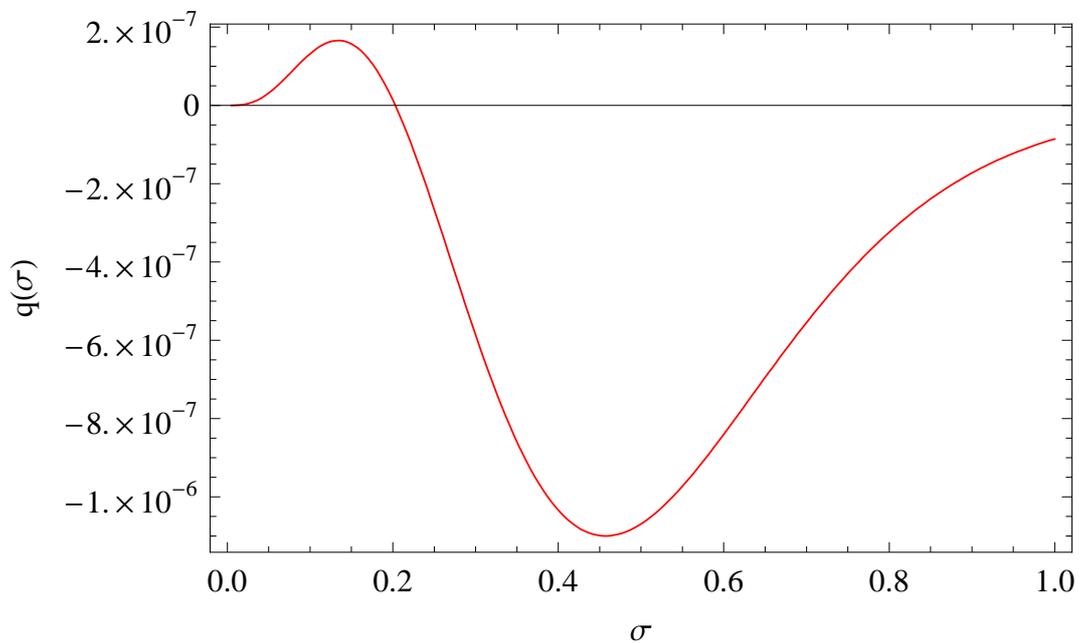}
\end{center}
\caption{[Color online] The three-loop function $q(\sigma)$ in terms of the parameter $\sigma = M_{\pi}/2\pi T$ for $N$=2.}
\label{figure3}
\end{figure}

Finally, the quantity $j$ in Eq.~(\ref{freeED}) is defined by
\begin{equation}
\label{definitionIj}
I = \frac{1}{{\pi}^2} g j \, ,
\end{equation}
where $I$ contains the contributions from the three-loop graphs 8A-C and reads
\begin{eqnarray}
\label{FunctionI}
I & = & \mbox{$ \frac{1}{48}$} (N-1)(N-3) M^4_{\pi} {\bar J}_1
- \mbox{$ \frac{1}{4}$} (N-1)(N-2) {\bar J}_2 \\
& & - \mbox{$ \frac{1}{16}$} (N-1)(N-3)^2 M^4_{\pi} (g_1)^2 g_2
+ \mbox{$ \frac{1}{48}$} (N-1)(N-3)(3N-7) M^2_{\pi} (g_1)^3 \, . \nonumber
\end{eqnarray}
The numerical evaluation of the renormalized three-loop integrals ${\bar J}_1$ and ${\bar J}_2$ has been described in Ref.~\citep{GL89},
and a graph for the function $j$ has been provided there for the specific case $N$=4. However, here we are interested in arbitrary
$N \ge 2$ -- accordingly, we have to evaluate ${\bar J}_1$ and ${\bar J}_2$ individually. This quite involved task has been performed in
Ref.~\citep{Hof16b} for ${\bar J}_1$. The numerical evaluation of ${\bar J}_2$, along the lines described in section 3 and the appendix of
Ref.~\citep{GL89}, has been performed in the present study. In Fig.~\ref{figure2} we show the resulting graphs for the function
$q(\sigma )$,
\begin{eqnarray}
\label{Functionq}
T^8 q(\sigma) & = & \mbox{$ \frac{1}{48}$} (N-1)(N-3) M^4_{\pi} {\bar J}_1 - \mbox{$ \frac{1}{4}$} (N-1)(N-2) {\bar J}_2 \nonumber \\
& = & \mbox{$ \frac{1}{48}$} (N-1)(N-3) q_1(\sigma) \, T^8 - \mbox{$ \frac{1}{4}$} (N-1)(N-2) q_2(\sigma) \, T^8 \, ,
\end{eqnarray}
for the cases $N = \{ 3,4,5,6 \}$. For completeness, in Fig.~\ref{figure3}, we depict $q(\sigma )$ for the abelian case $N$=2. We also
have checked consistency with Ref.~\citep{GL89}: setting $N$=4 indeed leads to the curve for $j$ displayed in figure 2 of that
reference.\footnote{Note that the function $j$ contains all contributions of $I$ (see Eqs.~(\ref{definitionIj}) and (\ref{FunctionI})),
whereas the function $q$, Eq.~(\ref{Functionq}), only involves the first two terms of $I$.}

\section{Low-Temperature Series}
\label{LowTSeries}

The effective field theory representations that we consider in this section are valid at low temperatures and weak external fields.
This means that both $T$ and $M$ (i.e., $H_s$) have to be small with respect to a typical scale $\Lambda$ that defines the microscopic
system under consideration. Concretely one may choose
\begin{equation}
T, M, H_s \ \lesssim 0.2 \ \Lambda \, .
\end{equation}
In quantum chromodynamics, $\Lambda$ is the renormalization group invariant scale $\Lambda_{QCD} \approx 1 GeV$. In Heisenberg or XY
magnets, the underlying scale can be identified with the exchange integral $J \approx 1 meV$. Regarding QCD, the leading-order effective
constant $F$ and the underlying scale $\Lambda$ are connected by \citep{MG84}
\begin{equation}
\label{LambdaF}
\Lambda = 4 \pi F \, .
\end{equation}
In the subsequent discussion, we assume that this relation is also valid for $N \neq 4$. Note that Eq.~(\ref{LambdaF}) refers to $d$=3+1
-- a generalization to arbitrary dimensions has been given in \citep{GJMM16}.

\subsection{Pressure}

Let us now discuss the low-temperature representation for the pressure,
\begin{equation}
P = z_0 - z \, .
\end{equation}
In order to display the explicit structure of $T$-powers, we introduce dimensionless functions $h_i(\sigma)$ and $h(\sigma)$ as follows:
\begin{eqnarray}
\label{ThermalDimensionless}
& & g_0(\sigma) = T^4 h_0(\sigma) , \qquad g_1(\sigma) = T^2 h_1(\sigma) , \qquad g(\sigma) = T^8 h(\sigma) \, , \nonumber \\
& & g_2(\sigma) = h_2(\sigma) , \qquad g_3(\sigma) = \frac{h_3(\sigma)}{T^2} \, , \qquad \quad \sigma = \frac{M_{\pi}}{2 \pi T} \, .
\end{eqnarray}
For $N$=2 we then obtain
\begin{eqnarray}
\label{pressureAbelian}
& & P(T,H_s) = p_1(\tau) \,T^4 + p_2(\tau) \, T^6 + p_3(\tau) \, T^8 + {\cal O}(T^{10}) \, , \hspace{2.0cm} (N=2) \nonumber \\
& & \qquad p_1(\tau) = \mbox{$ \frac{1}{2}$} h_0(\sigma) \, , \nonumber \\
& & \qquad p_2(\tau) = \frac{1}{8 F^2 t^2} {h_1(\sigma)}^2 \, , \nonumber \\
& & \qquad p_3(\tau) = \frac{3(e_1 + e_2) + \mbox{$ \frac{1}{2}$} {\overline k} - \mbox{$ \frac{3}{256 \pi^2}$ }}{F^4 \tau^4}
\, {h_1(\sigma)}^2 + \frac{2(e_1 + e_2)}{F^4} \, h(\sigma) \nonumber \\ 
& & \hspace{2.4cm} - \frac{1}{\pi^2 F^4} j(\sigma) h(\sigma) \, ,
\end{eqnarray}
where the renormalized coupling constant ${\overline k}$ is defined in appendix A of Ref.~\cite{Hof16b}.

In the nonabelian case ($N \ge 3$) we have
\begin{eqnarray}
\label{pressureNonAbelian}
& & P(T,H_s) = {\tilde p}_1(\tau) \, T^4 + {\tilde p}_2(\tau) \, T^6 + {\tilde p}_{3a}(\tau) \, T^8 + {\tilde p}_{3b} \, T^8 \, \ln
\Big( \frac{\Lambda_P}{T} \Big) + {\cal O}(T^{10} \ln T) \, , \nonumber \\
& & \quad {\tilde p}_1(\tau) = \mbox{$ \frac{1}{2}$} (N-1) h_0(\sigma) \, , \nonumber \\
& & \quad {\tilde p}_2(\tau) = - \frac{(N-1)(N-3)}{8 F^2 t^2} {h_1(\sigma)}^2 \, , \nonumber \\
& & \quad {\tilde p}_{3a}(\tau) = \frac{(N-1)}{32 \pi^2 F^4 \tau^4} \, \Big\{ (N+1)({\tilde \gamma_1} {\overline e_1} + {\tilde \gamma_2}
{\overline e_2}) + {\tilde \gamma_4} {\overline k_2} - {\tilde \gamma_3} {\overline k_1}
- \frac{3N^2 + 32N - 67}{24} \Big\} \nonumber \\
& & \hspace{2.4cm} \times {h_1(\sigma)}^2 + \frac{N-1}{32 \pi^2 F^4} \, \Big\{ 2 {\tilde \gamma_1} {\overline e_1} + N {\tilde \gamma_2}
{\overline e_2} - 2(N-2) \Big\} \, {\tilde h}(\sigma) \nonumber \\
& & \hspace{2.4cm} - \frac{1}{F^4} \, \Big\{ \frac{\pi^2}{675} {\tilde j(\sigma)} + \frac{j(\sigma) {\tilde h}(\sigma)}{\pi^2} \Big\} \, ,
\nonumber \\
& & \quad {\tilde p}_{3b}(\tau) = \frac{(N-1) (N-2) \pi^2}{6480 F^4} \, .
\end{eqnarray}
The renormalized NLO effective coupling constants ${\overline e_1}, {\overline e_2},{\overline k_1},{\overline k_2}$, the quantities
${\tilde \gamma_1}$, ${\tilde \gamma_2}$, ${\tilde \gamma_3}$, ${\tilde \gamma_4}$, the functions ${\tilde h}(\tau)$ and
${\tilde j}(\tau)$, as well as the scale $\Lambda_P$, are defined in the appendix.\footnote{Note that the two-loop contribution, both in
the abelian [$p_2(\tau)$] and the nonabelian [${\tilde p}_2(\tau)$] case, involves the ratio $t=T/M$, rather than $\tau=T/M_{\pi}$. The two
masses are related by Eqs.~(\ref{GBMass}) and (\ref{renMass}).} We have checked that for $N$=4 the coefficients ${\tilde p}_i(\tau)$
coincide with the known results for quantum chromodynamics at low temperatures, given explicitly in Ref.~\citep{GL89}.

The Goldstone-boson interaction comes into play beyond the free Bose gas term of order $T^4$. We define three classes of systems,
according to the sign of $p_2(\tau)$ and ${\tilde p}_2(\tau)$ that are associated with the dominant (two-loop) correction of order $T^6$.
In QCD, or for any other system with $N \ge 4$, the prefactor is negative and we are dealing with an attractive interaction among the
Goldstone bosons. On the other hand, in the quantum XY model (or any other realization of $N$=2), the prefactor is positive, signaling a
repulsion among the magnons. Finally, for the antiferromagnet ($N$=3) the two-loop term vanishes.

Interestingly, up to two-loop order, the low-temperature representation for the pressure exclusively involves powers of $T^2$. However, at
the three-loop level, logarithmic contributions of the form $T^8 \ln T$ show up in the {\it nonabelian} case. In fact, the three-loop
function $q(\sigma)$, Eq.~(\ref{Functionq}),
\begin{displaymath}
\label{q1q2}
T^8 q(\sigma) = \mbox{$ \frac{1}{48}$} (N-1)(N-3) q_1(\sigma) \, T^8 - \mbox{$ \frac{1}{4}$} (N-1)(N-2) q_2(\sigma) \, T^8 \, ,
\end{displaymath}
dictates whether or not such logarithms occur. The crucial piece is $q_2(\sigma)$, as it depends logarithmically on the parameter $\sigma$
or -- equivalently -- on $\tau = 1/(2\pi \sigma)$. In the abelian case, only $q_1(\sigma)$ is relevant: this function does not depend
logarithmically on $\sigma$ and one concludes that the three-loop contribution in the pressure then only involves a simple power $T^8$ --
no additional contribution $T^8 \ln T$ arises in the abelian case. The explicit steps that lead to the representations
(\ref{pressureAbelian}) and (\ref{pressureNonAbelian}) for the pressure can be found in the appendix.

Remarkably, for $N \ge 3$, in the limit $H_s, M_{\pi} \to 0$ only the logarithmic contribution that involves the scale $\Lambda_P$ survives
-- the other terms ($\propto T^8$ and $\propto T^6$) tend to zero. In the absence of an external field, the interaction contribution in
the pressure is thus dominated by the logarithm $T^8 \ln T$.

It should be pointed out that the presence or absence of logarithmic terms in the low-temperature representations is not an artifact of
our idealization concerning (pseudo-)Lorentz-invariance. The cateye graph 8C that defines the function $q_2(\sigma)$, exclusively involves
the leading-order piece ${\cal L}^2_{eff}$ that is strictly (pseudo-)Lorentz-invariant. The anisotropies of, e.g., the simple cubic lattice
start manifesting themselves at order ${\cal L}^4_{eff}$ and cannot affect the function $q(\sigma)$.

For the subsequent plots that illustrate the strength of the Goldstone-boson interaction, we need to know the numerical values of the
renormalized NLO effective constants ${\bar e}_1, {\bar e}_2, {\bar k}_1$ and ${\bar k}_2$. As described, e.g., in Ref.~\citep{GL84},
these (dimensionless) constants are of order one (see also appendix C of Ref.~\citep{Hof16b}),\footnote{It should be noted that we follow
the standard convention of chiral perturbation theory, Eq.~(\ref{LECstandardConvention}), where a factor of $1/32 \pi^2$ is introduced in
the relation between renormalized and unrenormalized NLO effective constants. In Ref.~\citep{Hof16b} we used a different convention (no
factor of $1/32 \pi^2$).}
\begin{equation}
\label{estimateNLO}
{\bar e}_1, {\bar e}_2, {\bar k}_1, {\bar k}_2 \ \approx \, 1 \, .
\end{equation}
In quantum chromodynamics, the values of the NLO effective constants (both their magnitude and sign) are known quite accurately. Those
relevant for our analysis are \citep{CGL01,BE14}\footnote{As we discuss in the appendix, the quantities ${\bar l}_1, \dots, {\bar l}_4$
have been evaluated at the scale $\mu = {\hat M}_{\pi}$.}
\begin{equation}
\label{valuesNLOEFCs}
{\bar l}_1 = -0.36 \pm 0.6 \, , \quad {\bar l}_2 = 4.31 \pm 0.1 \, , \quad{\bar l}_3 = 3.0 \pm 0.8 \, , \quad{\bar l}_4 = 4.39 \pm 0.2 \, .
\end{equation}
Notice that in our notation we have
\begin{equation}
\label{conversionNLOEFCs}
{\bar e}_1 = {\bar l}_1 \, , \qquad {\bar e}_2 = {\bar l}_2 \, , \qquad {\bar k}_1 = {\bar l}_4 \, , \qquad
{\bar k}_2 = {\bar l}_3 + {\bar l}_4 \, .
\end{equation}
Whereas the magnitude of ${\bar e}_1, {\bar e}_2, {\bar k}_1, {\bar k}_2$ can be estimated for arbitrary $N$, the sign of these effective
constants -- except for $N$=4 -- remains open: it depends on the specific realization of a given $N$ and must be determined for each
system individually.

One way to obtain the accurate numerical values and the signs of these NLO effective constants in connection with quantum magnets, would
be to compare the effective calculation with the analogous microscopic analysis -- unfortunately, such two- or three-loop results in the
parameter region of interest (low temperatures, weak external fields) appear to be unavailable. Another possibility to determine NLO
effective constants would be through numerical simulation of the underlying model, or by experiment. However, Monte Carlo simulations in
the relevant parameter region, or experimental results seem to be lacking. This then means that we have to rely our discussion on the
above estimates (except for QCD).
\begin{figure}
\includegraphics[width=13.5cm]{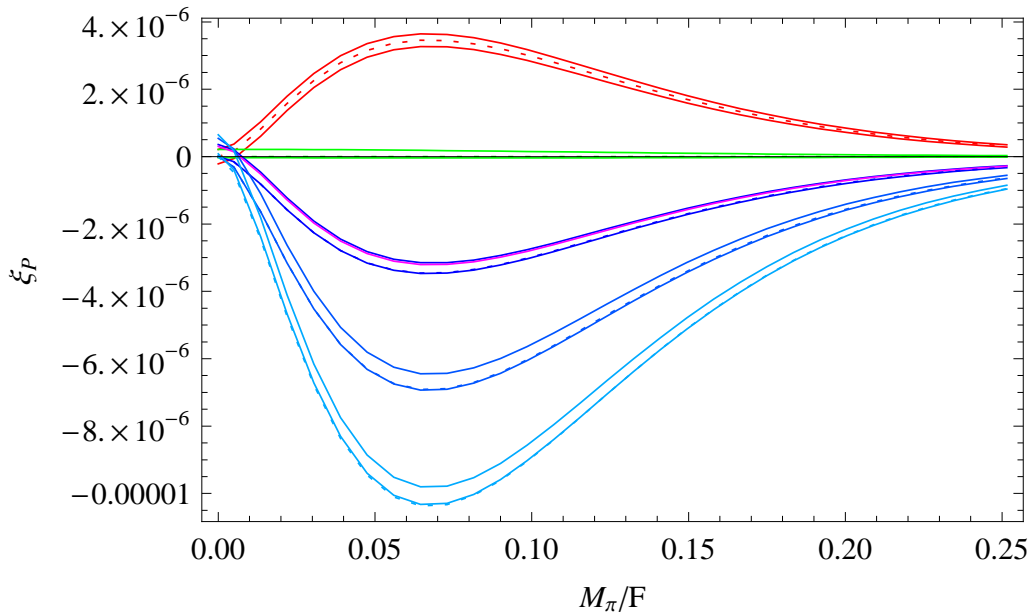}
\caption{[Color online] Goldstone-boson interaction manifesting itself in the pressure of $d$=(3+1) systems characterized by a
spontaneously broken symmetry O($N$) $\to$ O($N$-1) for $N$= \{2,3,4,5,6\} from top to bottom in the figure: two-loop contribution (dashed
curves) and sum of two-loop contribution and three-loop correction (continuous curves). The temperature is $T/F = 0.04$.}
\label{figure4}
\end{figure}

Let us now investigate the nature of the Goldstone boson interaction in presence of an external field $H_s$. In order to measure strength
and sign of the interaction in the pressure, in Fig.~\ref{figure4} we depict the quantity
\begin{equation}
\label{intRatioP}
\xi_P(T,H_s) = \frac{P^{[6]}_{int}(T,H_s) + P^{[8]}_{int}(T,H_s)}{P_{Bose}(T,H_s)}
\end{equation}
for the temperature $T/F = 0.04$ and $N=\{2,3,4,5,6 \}$. The dashed curves show the two-loop contribution ($P^{[6]}_{int}$), while the
continuous curves refer to the sum of the two-loop and the three-loop correction ($P^{[6]}_{int} + P^{[8]}_{int}$). Since the signs of the
effective constants ${\bar e}_1, {\bar e}_2, {\bar k}_1, {\bar k}_2$, in general, are unknown, we have scanned each of these couplings in
the interval
\begin{equation}
\label{scan}
\{ {\bar e}_1, {\bar e}_2, {\bar k}_1, {\bar k}_2 \} \ \subset \ [ -5, 5 ] \, ,
\end{equation}
in order to draw the three-loop correction. Note that these numbers correspond to the scale $\mu = {\hat M}_{\pi}$ -- according to
Eq.~(\ref{LECmassDependenceH}), the values of the effective constants ${\bar e}_1, {\bar e}_2, {\bar k}_1, {\bar k}_2$ depend on $M_{\pi}$,
i.e., on the external field $H_s$. Of course, this dependence has been taken into account in the figures. On the other hand, the
quantities ${\tilde \gamma_1}$, ${\tilde \gamma_2}$, ${\tilde \gamma_3}$ and ${\tilde \gamma_4}$ obey Eqs.~(\ref{consistencyM}) and
(\ref{consistencyAB}).

In Fig.~\ref{figure4} and all subsequent figures, we depict the two extreme situations for $\xi_P(T,H_s)$: the maximal and the minimal
three-loop corrections that we obtain from these scans -- this corresponds to an upper and lower bound for the three-loop contribution. In
the case of QCD, where both sign and magnitude of the NLO low-energy constants and also the ${\tilde \gamma_i}$'s are known, we have
highlighted the corresponding curve in magenta -- it perfectly lies between the lower and upper estimates.

\begin{figure}

\begin{center}

\includegraphics[width=13.5cm]{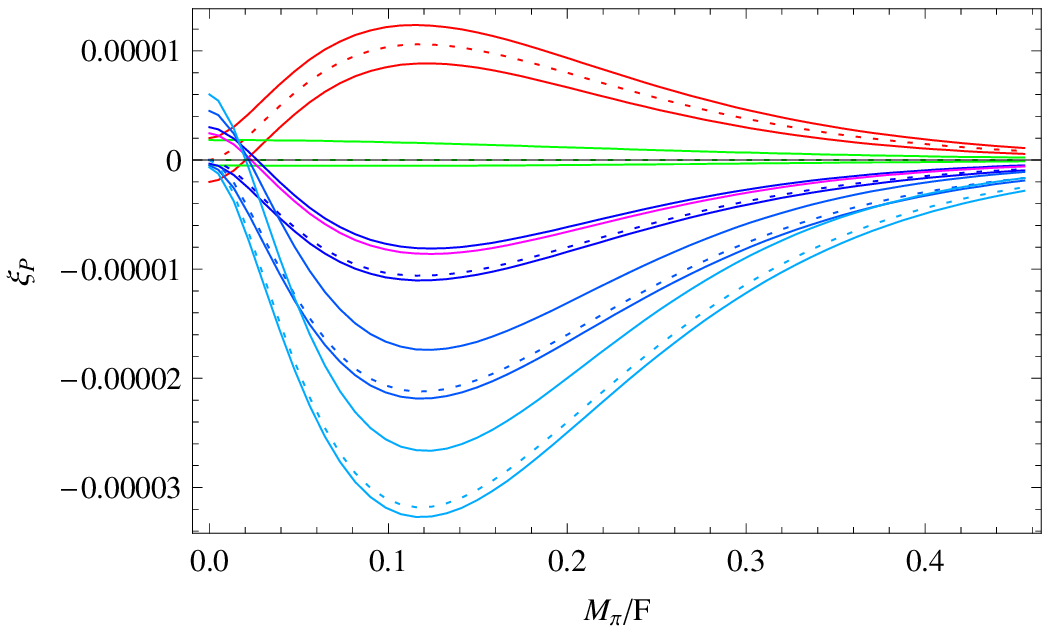} \\

\vspace{2mm}

\includegraphics[width=13.5cm]{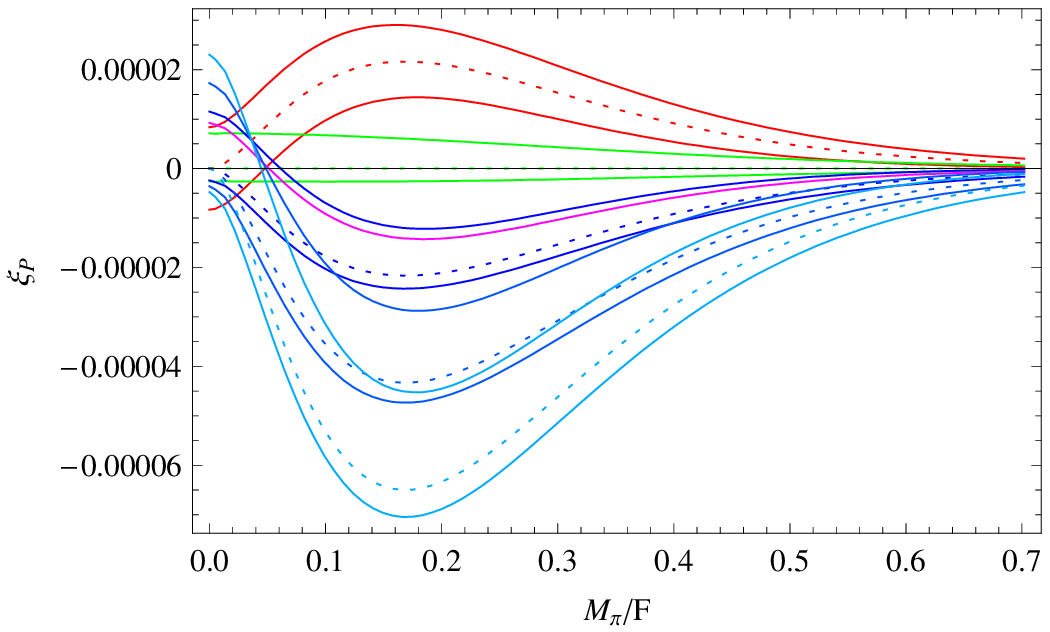} \\

\end{center}
\caption{[Color online] Goldstone-boson interaction manifesting itself in the pressure of $d$=(3+1) systems characterized by a
spontaneously broken symmetry O($N$) $\to$ O($N$-1) for $N$= \{2,3,4,5,6\} from top to bottom in the figure: two-loop contribution (dashed
curves) and sum of two-loop contribution and three-loop correction (continuous curves). The temperatures are $T/F = 0.07, 0.10$ (top,
bottom).}
\label{figure5}
\end{figure}

In QCD, the sign of the three-loop correction is positive and thus weakens the attractive two-loop contribution in the pressure. In
particular, in the chiral limit ($M_{\pi}, H_s \to 0$) the interaction among the pions becomes repulsive.\footnote{The {\it chiral limit}
in QCD refers to the fictitious world where the quark masses, and therefore the pion masses, are set to zero. In analogy to QCD, here we
also refer to $H_s \to 0$ as "chiral limit" when $N \neq 4$.} One notices that the three-loop correction for $N \neq 4$ in most scenarios
also tends to be positive as in QCD. Still, depending on the actual values of the NLO LEC's and the ${\tilde \gamma_i}$'s, the three-loop
correction in the pressure may be negative -- in particular, the Goldstone-boson interaction in the pressure may be attractive if the
external field is switched off. The three-loop corrections are quite small, but they grow when the temperature is raised, as we
illustrate in Fig.~\ref{figure5}.

\begin{figure}
\includegraphics[width=13.5cm]{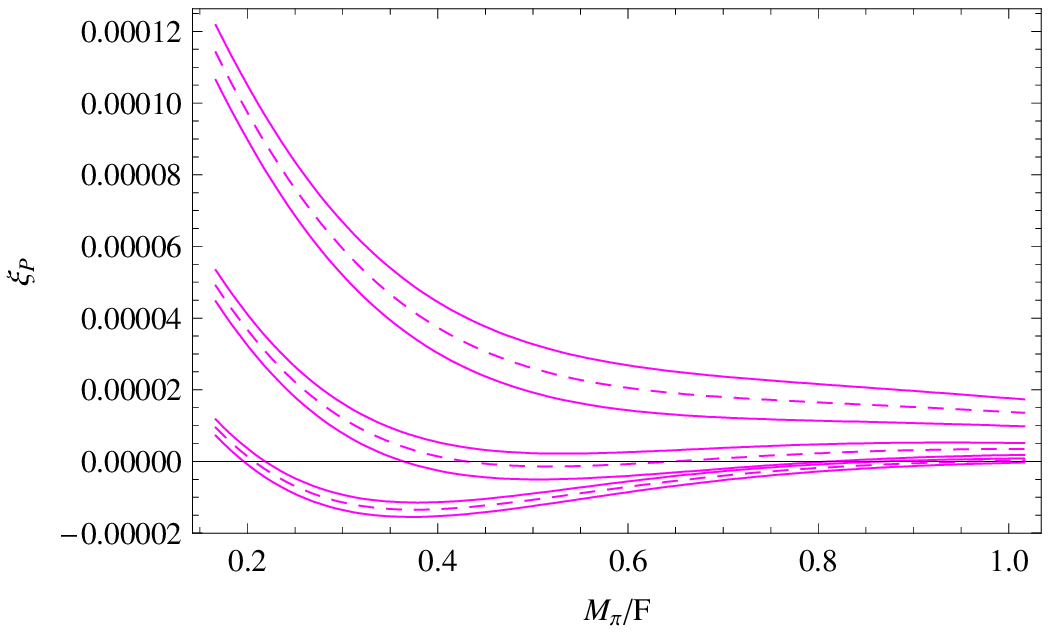}
\caption{Quantum chromodynamics ($N$=4): Manifestation of the pion-pion interaction in the pressure. The curves refer to $T/F = \{ 0.17,
0.20, 0.23 \}$ from bottom to top in the figure. The dashed curves correspond to the sum of the two- and three-loop correction. The error
bars reflect the uncertainty in the values of the NLO effective constants ${\bar e}_1, {\bar e}_2, {\bar k}_1, {\bar k}_2$ -- see
Eqs.~(\ref{valuesNLOEFCs}) and (\ref{conversionNLOEFCs}).}
\label{figure6}
\end{figure}

In QCD at low temperatures, according to Fig.~\ref{figure4}, the interaction among the pions is mainly attractive -- only for very small
ratios $M_{\pi}/F$ is becomes repulsive. However, as shown in Fig.~\ref{figure6}, the interaction eventually becomes repulsive in the
whole parameter regime $M_{\pi}/F$, provided that the temperature exceeds $T/F \approx 0.20$: the (positive) three-loop corrections then
become quite large. It should be pointed out that the interaction in the physical region -- where the pion mass is fixed at $M_{\pi}
\approx 139.6 MeV$ -- always is repulsive at low temperatures, and actually very weak: $\xi_P(T,M_{\pi}) \approx 10^{-6}$ for $T/F = 0.20$.

Unfortunately for systems with $N \neq 4$, as a consequence of the unknown signs of the NLO effective constants, the three-loop correction
in the pressure may be positive or negative, as witnessed by the previous figures. Accordingly, the sum of the two-loop and three-loop
contribution at more elevated temperatures may become repulsive in the entire parameter regime as in QCD. Again, to make concrete
statements on the nature of the interaction, the values of ${\bar e}_1, {\bar e}_2, {\bar k}_1, {\bar k}_2$ and the ${\tilde \gamma_i}$'s
not fixed by Eqs.~(\ref{consistencyM}) and (\ref{consistencyAB}), should be determined for each specific system -- for instance by Monte
Carlo simulations.

An important remark is finally in order here. Regarding the interaction one must distinguish between effects that already happen at zero
temperature and those that occur at finite temperature. The fact that the physical Goldstone boson mass $M_{\pi}$ and the bare Goldstone
boson mass $M$ are different, belongs to the first category: the corrections to the dominant contribution $M^2=\Sigma_s H_s/F^2$ in
$M^2_{\pi}$, Eq.~(\ref{renMass}), result from the interaction at $T$=0. Note that it is $M_{\pi}$ -- and not $M$ -- that is relevant in all
our figures where we show the interaction contribution at {\it finite} temperature parametrized by $M_{\pi}/F$. The interpretation of the
curves in our figures is thus the following: we begin at $T$=0 where the external field is switched on. Then we keep the field fixed, but
go to finite temperature -- the question then is whether switching on the temperature leads to an attraction or repulsion in the pressure.

\subsection{Order Parameter}

We now consider the order parameter, defined by
\begin{equation}
\Sigma_s(T,H_s) = - \frac{\partial z}{\partial H_s} \, .
\end{equation}
Depending on context, this is the spontaneous magnetization ($N$=2), the staggered magnetization ($N=2,3$), or the quark condensate
($N$=4). With the representation of the free energy density, for $N$=2 we obtain
\begin{eqnarray}
\label{OPabelian}
& & \Sigma_s(T,H_s) = \Sigma_s(0,H_s) + \sigma_1(\tau) T^2 + \sigma_2(\tau) T^4 + \sigma_3(\tau) T^6 +{\cal O}(T^8) \, ,
\nonumber \\
& & \qquad \sigma_1(\tau) = -\frac{\Sigma_s {\hat b}}{2 F^2} h_1(\sigma) \, , \nonumber \\
& & \qquad \sigma_2(\tau) = \frac{\Sigma_s}{8 F^4} \Big\{ {h_1(\sigma)}^2 -\frac{2 {\hat b}}{t^2}\, h_1(\sigma) h_2(\sigma) \Big\} \, ,
\nonumber \\
& & \qquad \sigma_3(\tau) = \frac{2 \Sigma_s}{t^2 F^6} \Big\{ 3(e_1 + e_2) + \mbox{$ \frac{1}{2}$} {\overline k} -
\mbox{$ \frac{3}{256 \pi^2}$} \Big\} \Big\{ {h_1(\sigma)}^2 - \frac{\hat b}{t^2} \, h_1(\sigma) h_2(\sigma) \Big\} \nonumber \\
& & \hspace{2.2cm} - \frac{3 \Sigma_s {\hat b}}{F^6} \Big\{ 2(e_1 + e_2) - \frac{1}{\pi^2}\, j(\sigma) \Big\}
\Big\{ h_0(\sigma) h_1(\sigma) + \frac{{h_1(\sigma)}^2 + h_0(\sigma) h_2(\sigma) }{\tau^2} \Big\} \nonumber \\
& & \hspace{2.2cm} - \frac{3 \Sigma_s {\hat b}}{8 \pi^4 F^6 \sigma} \frac{\partial j(\sigma)}{\partial \sigma} \, \Big\{ {h_0(\sigma)}^2
+ \frac{ h_0(\sigma) h_1(\sigma) }{\tau^2} \Big\} - \frac{\Sigma_s}{64 \pi^2 F^6 t^2} \, {h_1(\sigma)}^2 \, .
\end{eqnarray}
In the nonabelian case $(N \ge 3)$ we have 
\begin{eqnarray}
\Sigma_s(T,H_s) & = & {\tilde \Sigma}_s(0,H_s) + {\tilde \sigma}_1(\tau) T^2 + {\tilde \sigma}_2(\tau) T^4
+ {\tilde \sigma}_{3a}(\tau) \, T^6 + {\tilde \sigma}_{3b} \, T^6 \, \ln \Big( \frac{\Lambda_{\Sigma}}{T} \Big) \nonumber \\
& & + {\cal O}(T^8 \ln T) \, ,
\end{eqnarray}
with the coefficients
\begin{eqnarray}
& & \ {\tilde \sigma}_1(\tau) = -\frac{(N-1) \Sigma_s {\hat b}}{2 F^2} h_1(\sigma)\, , \nonumber \\
& & \ {\tilde \sigma}_2(\tau) = -\frac{ (N-1)(N-3) \Sigma_s}{8 F^4} \Big\{ {h_1(\sigma)}^2 -
\frac{2 {\hat b}}{t^2}\, h_1(\sigma) h_2(\sigma) \Big\} \, ,
\end{eqnarray}

\begin{eqnarray}
& & \ {\tilde \sigma}_{3a}(\tau) = \frac{(N-1)\Sigma_s}{16 \pi^2 \tau^2 F^6} \Big\{ (N+1)({\tilde \gamma_1} {\overline e_1}
+ {\tilde \gamma_2} {\overline e_2}) + {\tilde \gamma_4} {\overline k_2} - {\tilde \gamma_3} {\overline k_1}
- \frac{3N^2 + 32N - 67}{24} \Big\} \nonumber \\
& & \hspace{1.0cm} \times \Big\{ {h_1(\sigma)}^2 - \frac{\hat b}{\tau^2} \, h_1(\sigma) h_2(\sigma) \Big\}
- \frac{(N-1)\Sigma_s}{32 \pi^2 F^6 \tau^2} \, \Big\{ (N+1)({\tilde \gamma_1} + {\tilde \gamma_2}) + {\tilde \gamma_4}
- {\tilde \gamma_3} \Big\} \nonumber \\
& & \hspace{1.0cm} \times {h_1(\sigma)}^2
 - \frac{3 (N-1) \Sigma_s {\hat b}}{32 \pi^2 F^6} \Big\{ 2{\tilde \gamma_1} {\overline e_1}
+ N {\tilde \gamma_2} {\overline e_2} - 2(N-2) \Big\} \nonumber \\
& & \hspace{1.0cm} \times \Big\{ \frac{1}{12} {\tilde h_0}(\sigma) + \frac{\pi^2}{45} {\tilde h_1}(\sigma)
+ {\tilde h_0}(\sigma) {\tilde h_1}(\sigma)
+ \frac{{h_1(\sigma)}^2 + h_0(\sigma) h_2(\sigma) }{\tau^2} \Big\} \nonumber \\
& & \hspace{1.0cm} - \frac{(N-1) \Sigma_s\tau^2}{32 \pi^2 F^6} \, \Big\{ 2 {\tilde \gamma_1} + N {\tilde \gamma_2} \Big\}
\Big\{ {\tilde h}(\sigma) + \frac{\pi^4}{45} \, \sigma^2 \Big\} \nonumber \\
& & \hspace{1.0cm} + \frac{3 \Sigma_s {\hat b}}{\pi^2 F^6} \, \Big\{ \frac{1}{12} {\tilde h_0}(\sigma)
+ \frac{\pi^2}{45} {\tilde h_1}(\sigma) + {\tilde h_0}(\sigma) {\tilde h_1}(\sigma) + \frac{{h_1(\sigma)}^2
+ h_0(\sigma) h_2(\sigma) }{\tau^2} \Big\} \, j(\sigma) \nonumber \\
& & \hspace{1.0cm} + \frac{\Sigma_s {\hat b}}{180 F^6} \, {\tilde j}(\sigma)
- \frac{\Sigma_s {\hat b}}{5400 F^6 \sigma} \Big\{ \frac{\partial {\tilde j}(\sigma)}{\partial \sigma} \, - \, 8 \pi^2 j_2 \sigma\Big\}
- \frac{\Sigma_s {\hat b}}{8 \pi^4 F^6 \sigma} \frac{\partial {\tilde j}(\sigma)}{\partial \sigma} \, {\tilde h}(\sigma) \, ,
\nonumber \\
& & \ {\tilde \sigma}_{3b}(\tau) = - \frac{(N-1) (N-2) \Sigma_s}{1728 F^6} \, .
\end{eqnarray}
The parameter ${\hat b}$ is
\begin{equation}
\label{defb}
{\hat b}(H_s) = \frac{\partial M^2_{\pi}}{\partial M^2} = 1 + \frac{1}{16 \pi^2} \, \Big\{ {\tilde \gamma_4}( 2 {\overline k_2} - 1)
- {\tilde \gamma_3}(2 {\overline k_1} - 1) \Big\} \frac{ M^2}{F^2} + {\cal O}(H_s^2) \, .
\end{equation}
The Goldstone boson interaction in the order parameter manifests itself at order $T^4$ (two loops) and $T^6$ (three loops). Again, in the
nonabelian case, logarithmic contributions of the form $T^6 \ln T$ occur at the three-loop level that involve the scale $\Lambda_{\Sigma}$
that is defined in Eq.~(\ref{scaleLamq}). Analogous to the pressure, the term $\propto T^6 \ln(\Lambda_{\Sigma}/T)$ survives in the limit
$H_s, M_{\pi} \to 0$, whereas the other nonabelian three-loop contribution ($\propto T^6$) vanishes if the external field is switched off.

\begin{figure}

\begin{center}

\includegraphics[width=10.5cm]{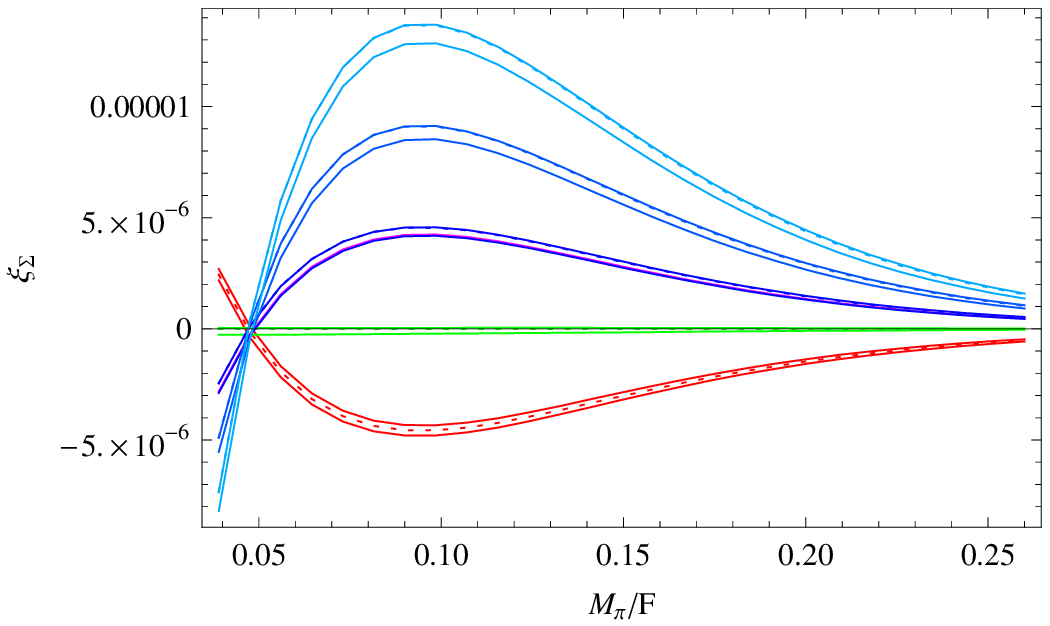} \\

\vspace{2mm}

\includegraphics[width=10.5cm]{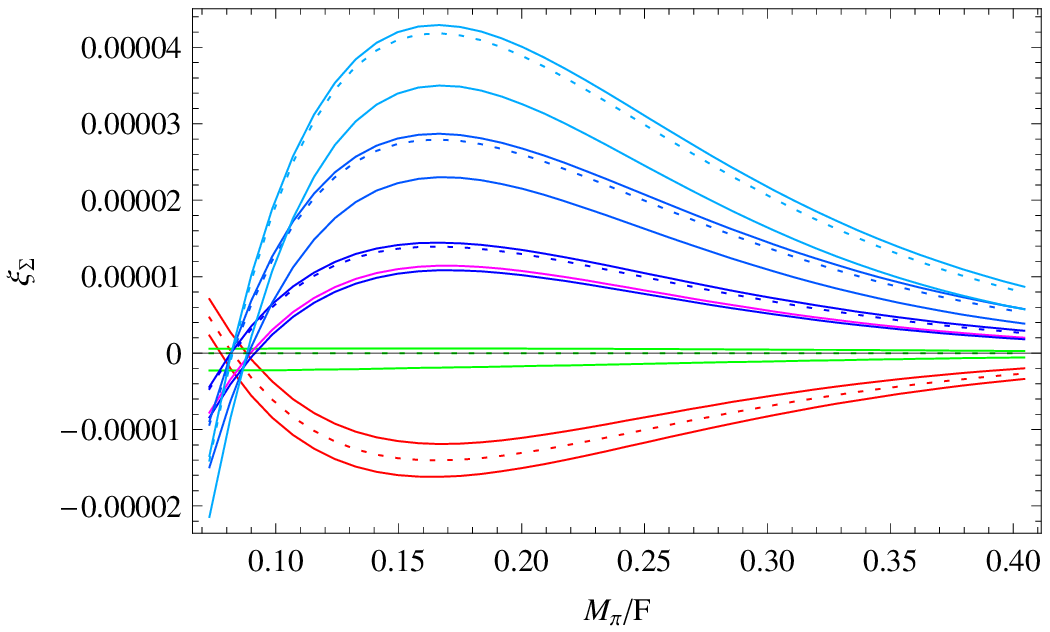} \\

\vspace{2mm}

\includegraphics[width=10.5cm]{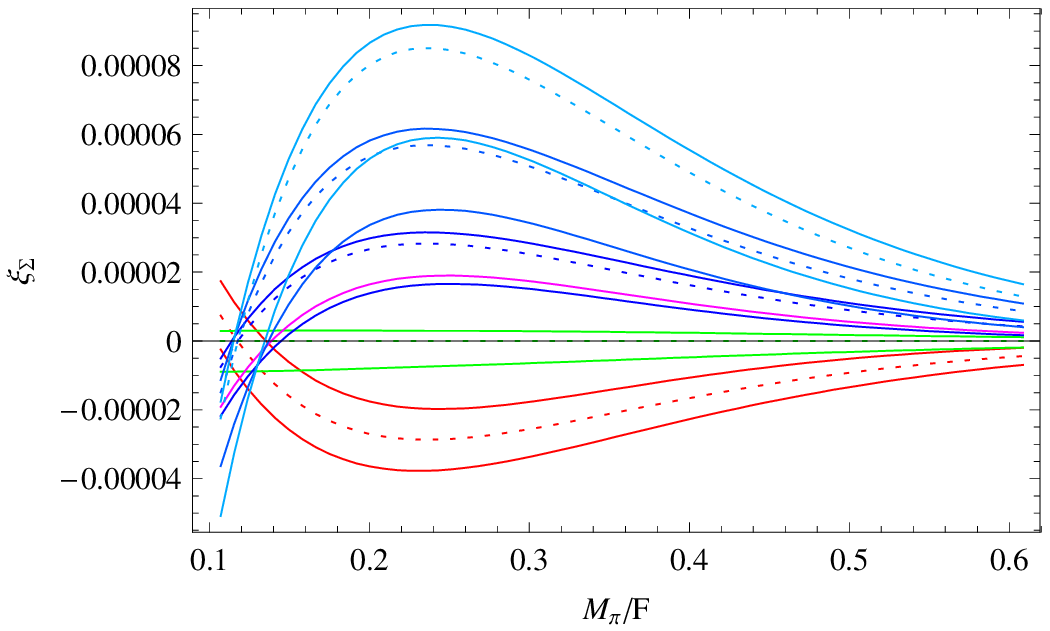} \\

\end{center}
\caption{[Color online] Temperature-dependent Goldstone-boson interaction manifesting itself in the order parameter of $d$=(3+1) systems
characterized by a spontaneously broken symmetry O($N$) $\to$ O($N$-1) for $N$= \{2,3,4,5,6\} from bottom to top in the figure (vertical
cut at $M_{\pi}/ F = 0.2$): two-loop contribution (dashed curves) and sum of two-loop contribution and three-loop correction
(continuous curves). The temperatures are $T/F = 0.04, 0.07, 0.10$ (top to bottom).}
\label{figure7}
\end{figure}

In order to measure strength and sign of the Goldstone boson interaction in the temperature-dependent part of order parameter, we consider
the dimensionless quantity
\begin{equation}
\xi_{\Sigma}(T,H_s) = \frac{{\Sigma}^{[4]}_{int}(T,H_s) + {\Sigma}^{[6]}_{int}(T,H_s)}{|{\Sigma}_{Bose}(T,H_s)|} \, ,
\end{equation}
where the numerator contains the sum of the two- and three-loop contribution. For $N$= \{2,3,4,5,6\}, in Fig.~\ref{figure7}, we depict the
above ratio $\xi_{\Sigma}(T,H_s)$ for the temperatures $T/F = 0.04, 0.07, 0.10$, respectively. The dashed curves correspond to the two-loop
contribution, while the continuous curves refer to the total correction (two-loop plus three-loop) that we obtained from the scans of NLO
low-energy effective constants as described before. Note that in the case of the antiferromagnet, the spin-wave interaction only starts
manifesting itself at three-loop order.

One notices that the three-loop corrections become quite large as the temperature rises. Depending on the specific realization of the
model, $\xi_{\Sigma}(T,H_s)$ may be positive. This is rather counterintuitive, because it means that the temperature-dependent interaction
enhances the order parameter when we go from zero to finite temperature while keeping the external field fixed. This happens for the
quantum XY model in weak external fields. For $N \ge 4$ and very low temperatures (see Fig.~\ref{figure7}), it basically occurs in the
entire region $M_{\pi}/F$ depicted, except for weak fields where $\xi_{\Sigma}(T,H_s)$ drops to negative values.

\begin{figure}
\includegraphics[width=13.5cm]{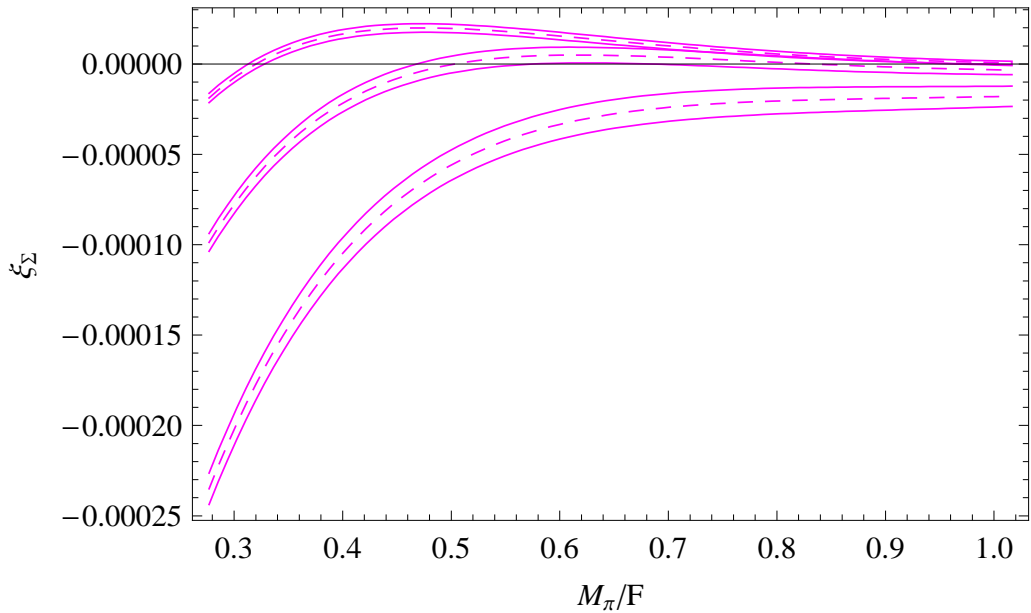}
\caption{Quantum chromodynamics ($N$=4): Manifestation of the pion-pion interaction in the quark condensate. The curves refer to
$T/F = \{ 0.17, 0.20, 0.23 \}$ from top to bottom in the figure. The dashed curves correspond to the sum of the two- and three-loop
correction. The error bars reflect the uncertainty in the values of the NLO effective constants ${\bar e}_1, {\bar e}_2, {\bar k}_1,
{\bar k}_2$ -- see Eqs.~(\ref{valuesNLOEFCs}) and (\ref{conversionNLOEFCs}).}
\label{figure8}
\end{figure}

In general, no definite conclusions regarding the sign of $\xi_{\Sigma}(T,H_s)$ are possible, because the signs of the NLO LEC's are
unknown. The exception is QCD where the corresponding curves in magenta indicate that $\xi_{\Sigma}(T,H_s)$ is positive at low temperatures
(except when approaching the chiral limit), but drops to negative values if the temperature is raised. Actually, if the temperature
exceeds $T/F \approx 0.20$, $\xi_{\Sigma}(T,H_s)$ attains negative values for any of the ratios $M_{\pi}/ F$ that we have considered. This
is illustrated in Fig.~\ref{figure8} . In the physical region ($M_{\pi} \approx 139.6 MeV$) the quantity $\xi_{\Sigma}(T,H_s)$ always is
negative at low temperatures -- but the effect of the interaction is very weak: $\xi_{\Sigma}(T,H_s) \approx - 10^{-6}$ for $T/F = 0.20$.
For QCD at the physical pion mass, there are thus no counterintuitive effects: if the temperature is raised while
$M_{\pi} \approx 139.6 MeV$ is kept fixed, the $T$-dependent interaction correction leads to a decrease of the quark condensate.

It should be pointed out that these two- and three-loop effects are very small. The properties of the order parameter are in fact
dominated by the one-loop contribution, which is negative in the entire parameter range for all $N$ that we have considered. The
respective term involves the quantity $- h_1$ that is negative (see Fig.~3 of Ref.~\citep{Hof16b}). As one expects, the order parameter
decreases if the temperature is raised while keeping the external field fixed.

\subsection{Susceptibility}

Finally, we consider the susceptibility,
\begin{equation}
\label{definitionSusceptibility}
\chi(T,H_s) = \frac{\partial \Sigma_s(T,H_s)}{\partial H_s} = \frac{\Sigma_s}{F^2} \, \frac{\partial \Sigma_s(T,M)}{\partial M^2} \, ,
\end{equation}
that describes the response of the order parameter to the applied external field. Both in the abelian and the nonabelian case, we obtain
\begin{eqnarray}
& & \chi(T,H_s) = \chi(0,H_s) + \chi_1(\tau) + \chi_2(\tau) T^2 + \chi_3(\tau) T^4
+ {\cal O}(T^6) \, , \nonumber \\
& & \qquad \chi_1(\tau) = \frac{(N-1) \Sigma^2_s}{2 F^4} \Big\{ {\hat b} \, h_2(\sigma) - \frac{{\hat b} \tau}{4 \pi} \,
\frac{\partial {\hat b}}{\partial \sigma} \, h_1(\sigma) \Big\} \, , \nonumber \\
& & \qquad \chi_2(\tau) = \frac{(N-1)(N-3)\Sigma^2_s}{8 F^6} \Bigg\{ \Big[ 2(1+{\hat b}) + \frac{{\hat b} \tau}{2 \pi t^2} \, 
\frac{\partial {\hat b}}{\partial \sigma} \Big] h_1(\sigma) h_2(\sigma) \nonumber \\
& & \hspace{2.6cm} - \frac{2 {\hat b}}{t^2} \Big( h_2(\sigma)^2 + h_1(\sigma) h_3(\sigma) \Big) \Bigg\} \, .
\end{eqnarray}
The explicit expression for the three-loop contribution $\chi_3(\tau)$ turns out to be lengthy. As it can trivially be obtained from the
representation of the order parameter, we do not list it here.

The temperature-dependent Goldstone boson interaction in the susceptibility shows up at order $T^2$ (two loops) and order $T^4$ (three
loops). In contrast to the pressure and the order parameter, the susceptibility becomes singular in the chiral limit $M_{\pi}, H_s \to 0$.
The dominant one-loop contribution diverges like
\begin{equation}
\lim_{H_s \to 0} \chi_1 \propto \frac{T}{\sqrt{H_s}} \propto \frac{T}{M} \, .
\end{equation}
The two-loop and three-loop contributions also become singular as the external field tends to zero. Therefore the decomposition of the
nonabelian three-loop contribution into two terms $\propto T^4$ and $\propto T^4 \ln(\Lambda_{\chi}/T)$ -- in order to analyze the
behavior of the system in the limit $M_{\pi}, H_s \to 0$ -- is purely "academic". Unlike for the pressure and the order parameter, in the
present case of the susceptibility, the three-loop terms diverge in the chiral limit and it makes no sense to introduce a scale
$\Lambda_{\chi}$. Remember that, concerning the pressure and the order parameter, the respective terms involving the scales $\Lambda_P$ and
$\Lambda_{\Sigma}$ remain finite in the chiral limit, while the other (non-logarithmic) contributions tend to zero.

The singular behavior of the one-loop contribution in the susceptibility has been reported for the $d$=3+1 quantum XY model in
Ref.~\citep{AHN80}. In quantum chromodynamics the divergence of the so-called disconnected chiral susceptibility has been reported first
in Ref.~\citep{SV96}, where two-loop corrections have been provided as well. We have to point out, however, that the analysis up to
three-loop order presented here, appears to be new for any $N = \{ 2,3,4\}$, to the best of our knowledge.

In the context of quantum spin models ($N=2,3$), the external field corresponds to the staggered field ${\vec H_s}=(H_s,0,0)$, and the
order parameter is identified with the staggered magnetization. Since ${\vec H_s}$ tends to reinforce the antiparallel arrangement of the
spins, one expects the order parameter to increase when the staggered field gets stronger -- the staggered susceptibility $\chi(T,H_s)$ is
therefore expected to be positive. Indeed, the leading (one-loop) contribution -- $\chi_1$ -- that refers to noninteracting spin waves, is
positive and dominates the behavior of the system. However, the manifestation of the spin-wave interaction -- or more general, Goldstone
boson interaction -- is more subtle, as we now discuss.

\begin{figure}

\begin{center}

\includegraphics[width=10.5cm]{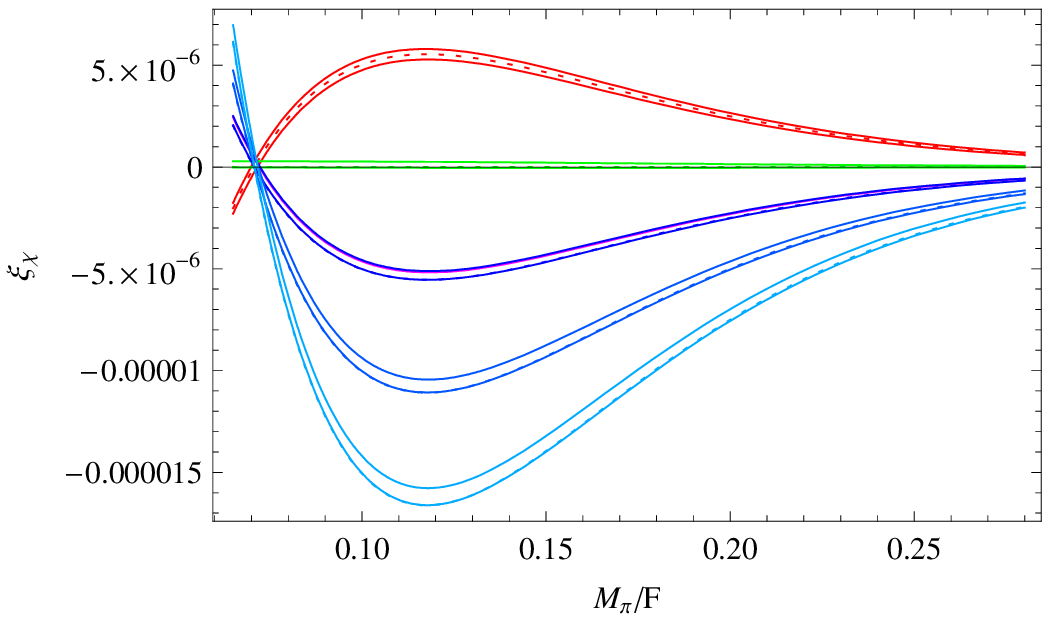} \\

\vspace{2mm}

\includegraphics[width=10.5cm]{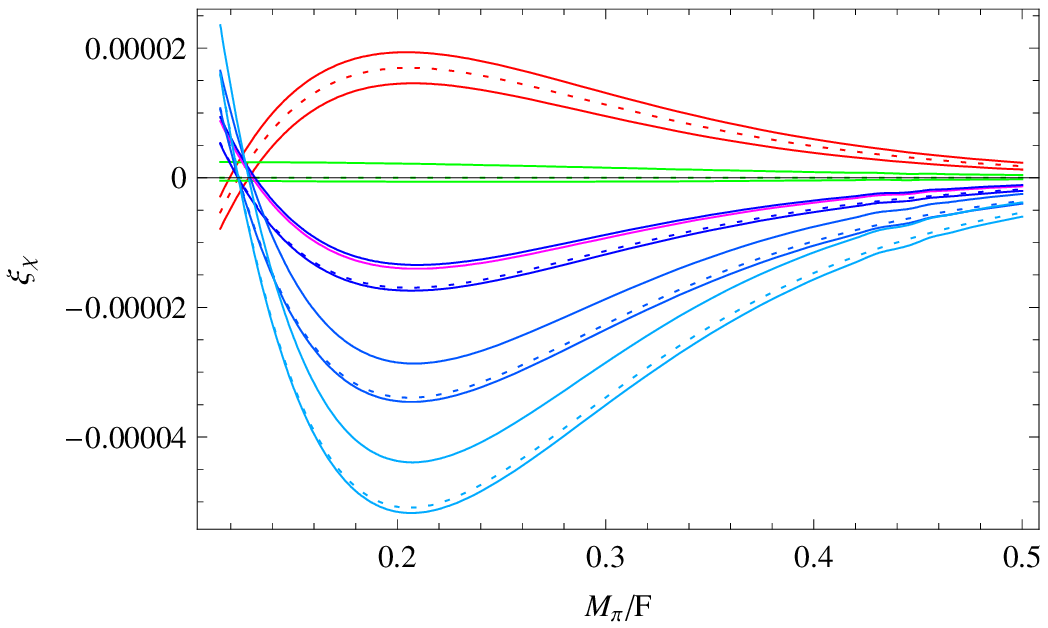} \\

\vspace{2mm}

\includegraphics[width=10.5cm]{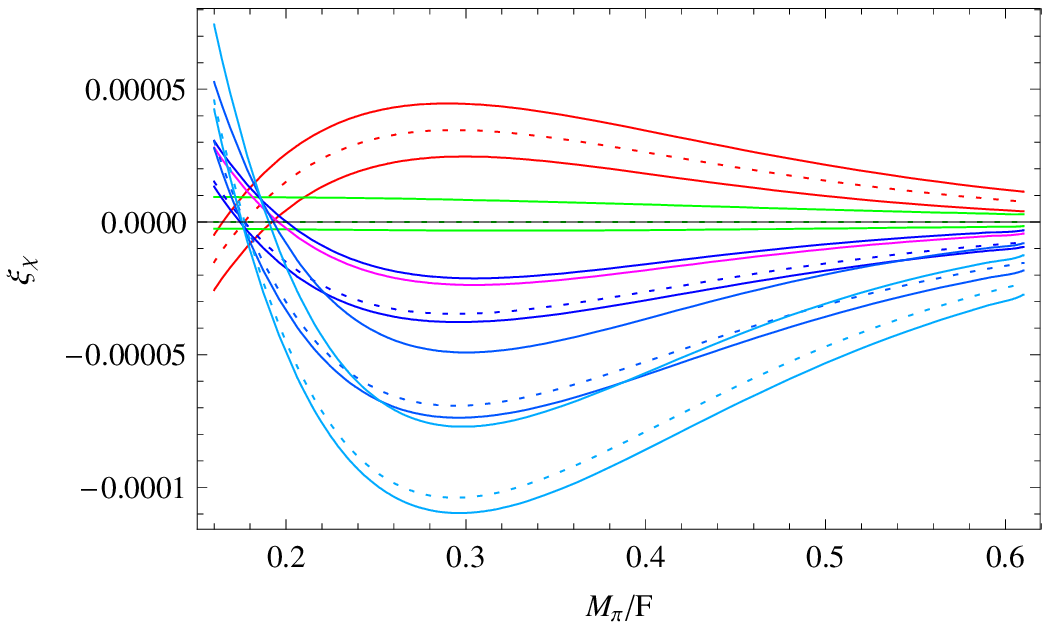}

\end{center}
\caption{[Color online] Temperature-dependent Goldstone-boson interaction manifesting itself in the susceptibility of $d$=(3+1) systems
characterized by a spontaneously broken symmetry O($N$) $\to$ O($N$-1) for $N$= \{2,3,4,5,6\} from top to bottom in the figure (vertical
cut at $M_{\pi}/F = 0.25$): two-loop contribution (dashed curves) and sum of two-loop contribution and three-loop correction
(continuous curves). The temperatures are $T/F = 0.04, 0.07, 0.10$ (top to bottom).}
\label{figure9}
\end{figure}

In order to measure the Goldstone boson interaction in the temperature-dependent part of the susceptibility, we define the dimensionless
quantity
\begin{equation}
\xi_{\chi}(T,H_s) = \frac{{\chi}^{[2]}_{int}(T,H_s) + {\chi}^{[4]}_{int}(T,H_s)}{{\chi}_{Bose}(T,H_s)} \, .
\end{equation}
In Fig.~\ref{figure9} we then plot the two-loop ($T^2$) and three-loop ($T^4$) contribution in $\xi_{\chi}(T,H_s)$ for three different
temperatures ($T/F = 0.04, 0.07, 0.10$) and $N=\{2,3,4,5,6 \}$. Note that we consider the behavior of the system in nonzero external
field, i.e., away from the limit $H_s \to 0$ where the susceptibility diverges. Much like in the pressure and the order parameter, the
spin-wave interaction in antiferromagnets only starts showing up at three-loop order.

\begin{figure}
\includegraphics[width=13.5cm]{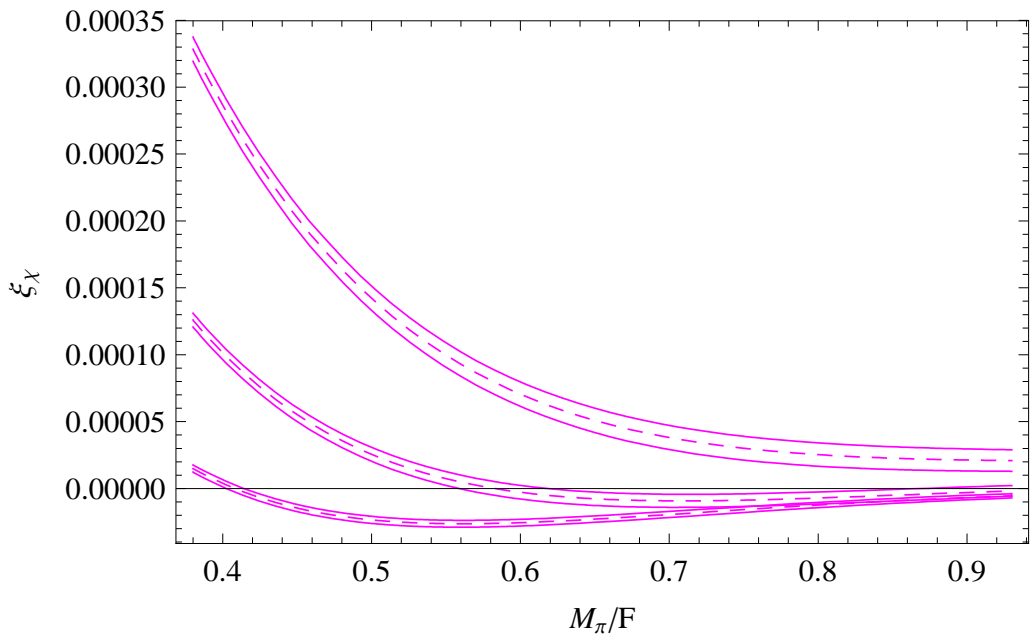}
\caption{Quantum chromodynamics ($N$=4): Manifestation of the pion-pion interaction in the susceptibility. The curves refer to
$T/F = \{ 0.17, 0.20, 0.23 \}$ from bottom to top in the figure. The dashed curves correspond to the sum of the two- and three-loop
correction. The error bars reflect the uncertainty in the values of the NLO effective constants ${\bar e}_1, {\bar e}_2, {\bar k}_1,
{\bar k}_2$ -- see Eqs.~(\ref{valuesNLOEFCs}) and (\ref{conversionNLOEFCs}).}
\label{figure10}
\end{figure}

The behavior of the $d$=3+1 quantum XY model is indeed peculiar, because the ratio $\xi_{\chi}(T,H_s)$ drops to negative values in weak
external fields. This behavior persists at more elevated temperatures (see Fig.~\ref{figure9}), and is contrary to all systems with
$N \ge 4$ where $\xi_{\chi}(T,H_s)$ tends to large positive values in weak external fields. If the temperature is very low ($T/F = 0.04$),
the parameter range where $\xi_{\chi}(T,H_s)$ is negative, is quite large for systems with $N \ge 4$. However, at more elevated
temperatures, the quantity $\xi_{\chi}(T,H_s)$ may reach positive values, since three-loop corrections become large. In particular, in QCD
where the NLO low-energy constants are known, $\xi_{\chi}(T,H_s)$ is positive in the entire parameter region $M_{\pi}/F$, as soon as the
temperature exceeds $T/F \approx 0.20$. This is illustrated in Fig.~\ref{figure10}. Also, in the physical region of QCD
($M_{\pi} \approx 139.6 MeV$), $\xi_{\chi}(T,H_s)$ is always positive and very weak at low temperatures: $\xi_{\chi}(T,H_s) \approx 10^{-6}$
for $T/F = 0.20$. We stress again that we are describing subtle two-loop and three-loop effects that manifest themselves in the
temperature-dependent interaction part of the susceptibility -- these effects are weak compared to the dominant noninteracting (one-loop)
contribution.

\section{Conclusions}
\label{conclusions}

The low-temperature behavior of $d$=3+1 (pseudo-)Lorentz-invariant systems with a spontaneously broken internal symmetry O($N$) $\to$
O($N$-1) can be derived systematically and straightforwardly up to three loops using effective Lagrangians. Important physical
realizations are the quantum XY model ($N$=2), the Heisenberg antiferromagnet ($N$=3), and two-flavor quantum chromodynamics ($N$=4).

From a conceptual point of view, the structure of the low-temperature series is of interest. In the nonabelian case, logarithmic
contributions of the form $({T^{2})}^n \ln T$ emerge at the three-loop level, whereas in the abelian case we are dealing with simple
powers of $T^2$. The coefficients accompanying the terms in the low-temperature expansion, depend in a nontrivial way on the ratio $H_s/T$
(where $H_s$ is an external field), and furthermore involve low-energy effective constants that parametrize the microscopic details of the
system.

One of our objectives was to investigate the impact of the Goldstone-boson interaction at finite temperature and nonzero external field
onto the thermodynamics of the system. In the pressure, the dominant two-loop contribution may be repulsive ($N$=2), zero ($N$=3), or
attractive ($N \ge 4$). Three-loop corrections -- depending on sign and magnitude of the NLO low-energy constants ${\bar e}_1, {\bar e}_2,
{\bar k}_1, {\bar k}_2$ -- may go into either direction. Overall, the three-loop corrections become substantial at more elevated
temperatures -- as a result, the Goldstone boson interaction in the pressure may become repulsive irrespective of $H_s$.

While the dominant one-loop contribution in the order parameter (susceptibility) is negative (positive), the sum of the two- and
three-loop correction may behave differently. These subtle effects, originating from the Goldstone boson interaction at finite
temperature, are particularly pronounced at low temperatures where the two-loop contribution dominates over the three-loop correction. The
$d$=3+1 quantum XY model is exceptional, since the temperature-dependent interaction contribution in the order parameter (susceptibility)
tends to large positive (negative) values in weak external fields. These somehow counterintuitive effects do not occur in QCD at the
physical pion mass.

The effective theory results regarding the Goldstone boson interaction, supersede well-cited results in the literature (e.g.,
Ref.~\citep{Ogu60}). Logarithmic terms in the low-temperature expansion of the order parameter of antiferromagnets have never been
reported outside the effective Lagrangian framework. Regarding quantum XY models, the analysis at low temperatures where the spin waves
dominate the physics of the system, has never been carried beyond one-loop order within the traditional microscopic perspective. In view
of these shortcomings in the literature, one concludes that the systematic effective Lagrangian method is a very valuable tool to address
the low-temperature properties of condensed matter systems.

As far as quantum chromodynamics is concerned, the subtle effects we observe in the pressure, quark condensate, and susceptibility, cannot
be "seen" in present lattice simulations. Our figures refer to very low temperatures ($T \lesssim  0.2 F$) and small (unphysical) pion
masses. While this parameter region is currently not accessible on the lattice, it is our hope that the subtle effects reported here may
be confirmed in future numerical studies.

It should be noted that our results are not restricted to quantum XY models, Heisenberg antiferromagnets, or two-flavor quantum
chromodynamics -- they also apply to other realizations of a specific $N$, provided the corresponding systems share the same symmetries.
While the low-temperature series then present the same structure -- powers of $T^2$ and logarithms $({T^{2})}^n \ln T$ -- the difference
concerns the specific numerical values of the effective constants that indeed depend on the system under consideration. This underlines
the universal nature of the effective field theory approach. What counts are the symmetries of the system -- the construction of the
effective Lagrangian then becomes a straightforward group-theoretical exercise and the corresponding effective field-theory predictions
are universal.

\section*{Acknowledgments}
The author thanks A.\ G\'omez Nicola, H.\ Leutwyler, A.\ Smilga and U.\ Wenger for correspondence.

\begin{appendix}

\section{Renormalization}
\label{appendix}

In this appendix we discuss some technical issues regarding renormalization. We start with the Goldstone boson mass $M_{\pi}$ that we
have defined in Eq.~(\ref{renMass}),\footnote{The explicit expression for the constant $c$ that involves next-to-next-to-leading order
effective constants is not needed here.}
\begin{equation}
\label{appendixRenMass}
M_{\pi}^2 = \frac{{\Sigma}_s H_s}{F^2} + \Big[ 2 (k_2 - k_1) + (N-3) \, \lambda \Big] \frac{({\Sigma}_s H_s)^2}{F^6}
+ c \frac{({\Sigma}_s H_s)^3}{F^{10}} + {\cal O} (H_s^4) \, ,
\end{equation}
and the parameters $a$ and $b$ defined in Eq.~(\ref{ConstAB}),
\begin{eqnarray}
\label{appendixConstAB}
a & = & - \frac{(N-1)(N-3)}{32{\pi}} \frac{{\Sigma}_s H_s}{F^4} + \frac{N-1}{4{\pi}} \frac{({\Sigma}_s H_s)^2}{F^8} \,
\Bigg\{ \Big[(N+1)(e_1 + e_2) + k_2 - k_1 \Big] \nonumber\\
& & - \frac{(N-1)^2}{2} \, \lambda - \frac{3N^2 + 32N - 67}{768 {\pi}^2} \Bigg\} \, , \nonumber \\ 
b & = & \frac{N-1}{\pi F^4} \, \Bigg\{ \Big[ 2 e_1 + N e_2 \Big] - \frac{5(N-2)}{3} \, \lambda - \frac{N-2}{16{\pi}^2} \Bigg\} \, .
\end{eqnarray}
These expressions involve unrenormalized (infinite) NLO effective constants: $e_1, e_2, k_1$ and $k_2$. Denoting these quantities as
$l_i \, (i=1,2,3,4)$, and adopting the standard convention of chiral perturbation theory (for details see, e.g., Ref.~\citep{GL84}),
\begin{equation}
\label{LECstandardConvention}
l_i = {\tilde \gamma_i} \Big( \lambda + \frac{1}{32 \pi^2} \, \overline l_i \Big) \, ,
\end{equation}
the singular expression $\lambda$ -- see Eq.~(\ref{lambda}) -- in $M_{\pi}, a$ and $b$ can be eliminated,
\begin{equation}
\label{constABfinite}
M_{\pi}^2 = \frac{{\Sigma}_s H_s}{F^2} + \frac{{\tilde \gamma_4} {\overline k_2} \! - \! {\tilde \gamma_3} {\overline k_1}}{16 {\pi}^2}
\, \frac{({\Sigma}_s H_s)^2}{F^6} + {\cal O} (H_s^3) \, ,
\end{equation}
\begin{eqnarray}
a & = & - \frac{(N-1)(N-3)}{32{\pi}} \frac{{\Sigma}_s H_s}{F^4} + \frac{N-1}{128 \pi^3} \frac{({\Sigma}_s H_s)^2}{F^8} \, \nonumber \\
& & \times \Bigg\{ (N+1)({\tilde \gamma_1} {\overline e_1} + {\tilde \gamma_2} {\overline e_2}) + {\tilde \gamma_4} {\overline k_2}
- {\tilde \gamma_3} {\overline k_1} - \frac{3N^2 + 32N - 67}{24} \Bigg\} \, , \nonumber \\
b & = & \frac{N-1}{32 \pi^3 F^4} \, \Bigg\{ 2 {\tilde \gamma_1} {\overline e_1} + N {\tilde \gamma_2} {\overline e_2} - 2(N-2) \Bigg\} \, .
\end{eqnarray}
Up to a factor of ${\tilde \gamma_i}/32 \pi^2$, the constants $\overline l_i$ are the so-called running coupling constants at the fixed
scale $\mu = {\hat M}_{\pi}$, where ${\hat M}_{\pi} \approx 139.6 MeV$ is the physical pion mass. The quantities ${\tilde \gamma_i}$ are
coefficients of order one. To avoid confusion, we explicitly provide our renormalization conventions:
\begin{eqnarray}
\label{ExplicitConventions}
& & e_1 = {\tilde \gamma_1} \Big( \lambda + \frac{1}{32 \pi^2} \, \overline e_1 \Big) \, , \qquad
e_2 = {\tilde \gamma_2} \Big( \lambda + \frac{1}{32 \pi^2} \, \overline e_2 \Big) \, , \qquad 
k_1 = {\tilde \gamma_3} \Big( \lambda + \frac{1}{32 \pi^2} \, \overline k_1 \Big) \, , \nonumber \\
& & k_2 = {\tilde \gamma_4} \Big( \lambda + \frac{1}{32 \pi^2} \, \overline k_2 \Big) \, .
\end{eqnarray}
The fact that the singular quantity $\lambda$ must drop out in the renormalized expressions $M_{\pi}, a, b$, implies various consistency
conditions among the coefficients ${\tilde \gamma_i}$. From the renormalization of $M_{\pi}^2$ one concludes
\begin{equation}
\label{consistencyM}
{\tilde \gamma_4} - {\tilde \gamma_3} = -\frac{(N-3)}{2} \, .
\end{equation}
Invoking the renormalization of the parameters $a$ and $b$ one obtains
\begin{eqnarray}
\label{consistencyAB}
& & {\tilde \gamma_1} = \frac{3N-10}{6} \, , \qquad {\tilde \gamma_2} = \frac{2}{3} \, , \qquad (N \neq 2) \, , \nonumber \\
& & {\tilde \gamma_1} + {\tilde \gamma_2} = 0 \, , \qquad (N = 2) \, . 
\end{eqnarray}
In QCD ($N$=4), according to Ref.~\citep{GL84}, we have ${\tilde \gamma_1}=\frac{1}{3}$ , ${\tilde \gamma_2}=\frac{2}{3}$,
${\tilde \gamma_4} - {\tilde \gamma_3} = -\frac{1}{2}$: indeed, the above consistency conditions are satisfied.

The renormalized NLO effective constants $\overline l_i$ depend on the pion mass $M_{\pi}$ as follows:
\begin{equation}
\label{LECmassDependence}
{\overline l_i} = - \ln {\Big( \frac{M_{\pi}}{{\hat M}_{\pi}} \Big)}^2 + {\hat l_i} \, .
\end{equation}
Here the $\hat l_i$ are pure numbers of order one. While the mass $M_{\pi}$ varies according to the strength of the field $H_s$, note that
the mass ${\hat M}_{\pi}$ corresponds to the physical (fixed) pion mass ${\hat M}_{\pi} = 139.6 MeV$. To have a general relation that is
also valid for $N \neq 4$, we rewrite Eq.~(\ref{LECmassDependence}) in the form
\begin{equation}
\label{LECmassDependenceH}
{\overline l_i} = - \ln {\Big( \frac{M_{\pi}/F}{{\hat M}_{\pi}/F} \Big)}^2 + {\hat l_i} \, ,
\end{equation}
where the masses are now given in relation to the leading-order effective constant $F$. In analogy to QCD, for $N \neq 4$ we choose the
ratio in question as
\begin{equation}
\frac{{\hat M}_{\pi}}{F} \equiv \frac{139.6 MeV}{86.8 MeV} \approx 1.61 \, .
\end{equation}
In connection with magnetic systems, ${\hat M}_{\pi}$ corresponds to a (fixed) external field according to Eq.~(\ref{renMass}). It should
be pointed out that the quantities $\overline l_i$ become singular in the chiral limit $M_{\pi} \to 0$ ($|{\vec H}_s| \to 0$). It so seems
that taking this limit in the free energy density -- and in all thermodynamic observables derived from there -- is problematic.

The crucial point, however, is the following: in the free energy density, Eq.~(\ref{freeED}), apart from the parameter $b$ that explodes
in the chiral limit, the function $j$ also becomes singular if $M_{\pi} \to 0$,\footnote{Note that the parameter $a$ tends to zero in the
chiral limit.}
\begin{equation}
\label{logj}
j(\tau) = \frac{5(N-1)(N-2)}{48} \, \ln \tau + {\hat j}(\tau) \, , \qquad \tau = \frac{T}{M_{\pi}} \, .
\end{equation}
Using the previous formulas, the parameter $b$ -- in the {\it nonabelian} case -- can be rewritten as
\begin{equation}
\label{bwithScale}
b = - \frac{5(N-1)(N-2)}{48{\pi}^3 F^4} \, \ln\frac{M_{\pi}}{\Lambda_b} \, , \qquad (N \neq 2) \, ,
\end{equation}
where the scale $\Lambda_b$ involves NLO effective constants,
\begin{equation}
\label{ScaleLambdaB}
\Lambda_b = {\hat M}_{\pi} \, \exp\Big[ \frac{6 {\tilde \gamma_1} {\hat e}_1 + 3 N {\tilde \gamma_2} {\hat e}_2}{10(N-2)}
- \frac{3}{5} \Big]\, , \qquad (N \neq 2) \, .
\end{equation}
The coefficient of the logarithm in $b$ is proportional to $(N-1)(N-2)$, much like the coefficient of the logarithm that appears in
$j(\tau)$. Hence the two logarithms can be merged into a single logarithm that remains finite if $M_{\pi} \to 0$. More concretely, the
relevant piece in the representation of the free energy density, Eq.~(\ref{freeED}), is
\begin{equation}
g \, \Big[ b - \frac{j}{{\pi}^3 F^4} \Big] = h \, \Big[ b - \frac{j}{{\pi}^3 F^4} \Big] \, T^8 \, .
\end{equation}
Since the limit $M_{\pi} \to 0$ plays a special role, we decompose the functions $h$ and $j$ as
\begin{equation}
\label{hjDecomp}
h(\tau) = \frac{\pi^4}{675} + {\tilde h}(\tau) \, , \qquad j(\tau) = \nu \, \ln \tau + j_1 + {\tilde j}(\tau) \, , \quad
\nu = \frac{5(N-1)(N-2)}{48} \, .
\end{equation}
For $M_{\pi} \to 0$ (and $T$ fixed), the remainders ${\tilde h}(\tau)$ and ${\tilde j}(\tau)$ tend to zero by construction. Using the
representations (\ref{bwithScale}) and (\ref{hjDecomp}), we end up with three contributions,
\begin{eqnarray} 
h \, \Big[ b - \frac{j}{{\pi}^3 F^4} \Big] & = & \frac{1}{32 \pi^2} \, \Big\{ 2 {\tilde \gamma_1} {\overline e_1} + N {\tilde \gamma_2}
{\overline e_2} - 2(N-2) \Big\} \, {\tilde h}(\sigma) \\
& & - \frac{1}{N-1} \, \Big\{ \frac{\pi^2}{675} {\tilde j(\sigma)} + \frac{j(\sigma) {\tilde h}(\sigma)}{\pi^2} \Big\}
+ \frac{(N-2) \pi^2}{6480} \, \ln \Big( \frac{\Lambda_P}{T} \Big) \, . \nonumber
\end{eqnarray}
In the chiral limit, only the last term survives. The scale $\Lambda_P$ is defined by
\begin{eqnarray}
\Lambda_P & = & {\hat M}_{\pi} \, \exp\Big[ \frac{6 {\tilde \gamma_1} {\hat e}_1 + 3 N {\tilde \gamma_2} {\hat e}_2}{10(N-2)} - \frac{3}{5}
- \frac{48 j_1}{5(N-1)(N-2)} \Big] \, , \quad (N \neq 2) \, , \nonumber \\
& = & \Lambda_b \exp\Big[ - \frac{48 j_1}{5(N-1)(N-2)} \Big] \, .
\end{eqnarray}
We have evaluated the coefficients $j_1$ following the procedure described in Ref.~\citep{GL89}. The results are provided in the second
column of Table \ref{table1} for $N = \{ 3, 4, 5, 6 \}$.

\begin{table}[h!]
\centering
\begin{tabular}{|c||c|c|c|}
\hline
N  &  $\nu$   & $j_1$    & $j_2$ \\
\hline
\hline
3  &  $5/24$  &  0.1122  &  0.0363  \\
\hline
4  &  $5/8$   &  0.3366  &  0.119   \\
\hline
5  &  $5/4$   &  0.6731  &  0.260   \\
\hline
6  &  $25/12$ &  1.122   &  0.472   \\
\hline
\end{tabular}
\caption{\it The coefficients $\nu$, $j_1$ and $j_2$ referring to the three-loop function $j(\tau)$.}
\label{table1}
\end{table}

Collecting results, the low-temperature expansion for the pressure $P = z_0-z$ takes the form
\begin{eqnarray}
\label{repPressure}
P & = & \mbox{$ \frac{1}{2}$} (N-1) h_0 T^4 - \frac{(N-1)(N-3)}{8 F^2 t^2} \, h_1^2 T^6 \nonumber \\
& & + \frac{N-1}{32 \pi^2 F^4 \tau^4} \, \Bigg\{ (N+1)({\tilde \gamma_1} {\overline e_1} + {\tilde \gamma_2}
{\overline e_2}) + {\tilde \gamma_4} {\overline k_2} - {\tilde \gamma_3} {\overline k_1}
- \frac{3N^2 + 32N - 67}{24} \Bigg\} h_1^2 \, T^8 \nonumber \\
& & + \frac{(N-1) (N-2) \pi^2}{6480 F^4} \, T^8 \, \ln \Big( \frac{\Lambda_P}{T} \Big) 
+ \frac{N-1}{32 \pi^2 F^4} \, \Bigg\{ 2 {\tilde \gamma_1} {\overline e_1} + N {\tilde \gamma_2} {\overline e_2} - 2(N-2) \Bigg\} \, 
{\tilde h} T^8 \nonumber \\
& & - \frac{1}{F^4} \, \Big\{ \frac{\pi^2}{675} {\tilde j} + \frac{j {\tilde h}}{\pi^2} \Big\} T^8 \, .
\end{eqnarray}
This representation is valid for the nonabelian case $N \ge 3$. In the abelian case, matters are much simpler because the singular piece
$\lambda$ in the parameter $b$, Eq.~(\ref{appendixConstAB}), does not occur. Here the combination $e_1 + e_2$ is finite -- in particular,
it does not explode in the chiral limit. Likewise, for $N$=2, the three-loop function $j$ tends to zero in the chiral limit and does not
become singular -- everything is consistent also in the abelian case, where no logarithmic contribution $T^8 \ln T$ in the free energy
density or pressure emerges.

In the low-temperature expansion of the order parameter, another scale -- ${\Lambda}_{\Sigma}$ -- enters at order $T^6 \ln T$. It is
related to the scale ${\Lambda}_P$ showing up in the pressure by 
\begin{equation}
\label{scaleLamq}
{\Lambda}_{\Sigma} = \Lambda_P \, \exp\Big(\frac{4 \pi^2}{15 \nu} j_2\Big) \, .
\end{equation}
The above coefficients $j_2$ occur in the expansion of the function ${\tilde j}(\tau)$,
\begin{equation}
{\tilde j}(\tau) = \frac{j_2}{\tau^2} + \frac{j_3}{\tau^3} + {\cal O}(\tau^{-4}) \, ,
\end{equation}
and are listed in the third column of Table \ref{table1} for $N = \{ 3, 4, 5, 6 \}$.

When evaluating the order parameter and the susceptibility on the basis of the representation for the pressure, Eq.~(\ref{repPressure}),
one should keep in mind that the renormalized NLO effective constants obey
\begin{equation}
\frac{\partial {\overline l_i}}{\partial M^2} = - \frac{1}{M^2} \, .
\end{equation}
Again, NLO effective constants that are not renormalized, do not depend on $M$ (or $|{\vec H}_s|$).

\end{appendix}

\end{document}